\documentclass[draftclsnofoot,onecolumn,11pt]{IEEEtran}
\usepackage[T1]{fontenc}
\normalsize

\ifCLASSINFOpdf
\else
\fi

\usepackage[cmex10]{amsmath}
\usepackage{amsfonts}
\usepackage{amssymb}
\usepackage{amsthm}
\usepackage{calrsfs}
\usepackage{color}
\usepackage{xcolor}
\usepackage{graphicx}
\usepackage{bm}

\usepackage{array}
\usepackage{algorithmic}
\usepackage{array}
\usepackage{url}
\usepackage{stmaryrd}
\newtheorem{theorem}{Theorem}

\newtheorem{remark}{Remark}

\newcommand{\cE}{\mathcal{E}}
\newcommand{\cB}{\mathcal{B}}

\newcommand{\bX}{\bm{X}}

\newcommand{\bV}{\bm{V}}

\newcommand{\bv}{\bm{v}}
\newcommand{\bx}{\bm{x}}
\newcommand{\bu}{\bm{u}}
\newcommand{\latpt}{{\bm{y}}}
\newcommand{\Latpt}{{\bm{Y}}}
\newcommand{\Vor}{\mathcal{V}}
\newcommand{\Bab}{\mathcal{B}}

\newcommand{\Intervalone}{\mathcal{I}}
\newcommand{\Intervaltwo}{\mathcal{J}}

\newcommand{\VInterval}{\mathcal{ J}}

\newcommand{\mynint}[1]{\left\llbracket #1 \right\rrbracket}

\newcommand{\vol}[1]{\mbox{vol}\left(#1\right)}
\newcommand{\prob}[1]{{Prob}\left\{#1\right\}}
\newcommand{\Rate}{\zeta}

\newcommand{\vzero}{\mathbf{0}}
\newcommand{\ent}[1]{{H}\left(#1\right)}

\newcommand{\conent}[2]{{H}\left(#1\mid#2\right)}

\newcommand{\heightbabai}{|v_{2,2}|}
\newcommand{\lengthbabai}{|v_{1,1}|}
\newcommand{\lenthree}{L_{W_1,Z_1,Z_2}(w_1,z_1,z_2)}
\newcommand{\lentwo}{L_{W_1,Z_1}(w_1,z_1)}
\newcommand{\lenone}{L_{W_1}(w_1)}
\newcommand{\Intthree}{\mathcal{I}_{W_1,Z_1,Z_2}(w_1,z_1,z_2)}
\newcommand{\Inttwo}{\mathcal{I}_{W_1,Z_1}(w_1,z_1)}
\newcommand{\Intone}{\mathcal{I}_{W_1}(w_1)}
\newcommand{\Intonezero}{\mathcal{I}_{W_1}(0)}
\newcommand{\Rthree}{\mathcal{R}_{W_1,Z_1,Z_2}(w_1,z_1,z_2)}
\newcommand{\Rtwo}{\mathcal{R}_{W_1,Z_1}(w_1,z_1)}
\newcommand{\Rone}{\mathcal{R}_{W_1}(w_1)}
\newcommand{\Ronezero}{\mathcal{R}_{W_1}(0)}
\newcommand{\Rblank}{\mathcal{R}}
\usepackage{tikz}
\usetikzlibrary{shadows,shapes,arrows} 
\usepackage[framemethod=tikz]{mdframed} 
\usepackage{pgfplots}
\usetikzlibrary{calc} 
\usepgfplotslibrary{fillbetween}
\usetikzlibrary{positioning}
\usepackage{ctable} 

\usepackage{balance}

\begin{document}
%
\title{Interactive Nearest Lattice Point Search in a Distributed Setting: Two Dimensions} 
%
%
%

\author{V.~A.~Vaishampayan
~and~M.~F.~Bollauf
\thanks{V. A. Vaishampayan is with the Department
of Electrical  Engineering, City University of New York, College of Staten Island, Staten Island,
NY 10314, USA e-mail: Vinay.Vaishampayan@csi.cuny.edu}
\thanks{M. ~F. ~Bollauf is with Simula UiB, Norway and was a visiting scholar at CUNY in 2016 e-mail maiara@simula.no}%
\thanks{This work was presented in part at the 2017 IEEE International Symposium on Information Theory, see \cite{VB:2017}}}

\markboth{IEEE Transactions on Communications}%
{Submitted paper}
%



\maketitle

\vspace{-0.5in}
\begin{abstract}
The  nearest lattice point problem in $\mathbb{R}^n$ is formulated in a distributed network with $n$ nodes. The objective is to minimize the probability that an incorrect lattice point is found, subject to a constraint on  inter-node communication. Algorithms with a single as well as an unbounded number of rounds of communication are considered for the case $n=2$. For the algorithm with a single round, expressions are derived for the error probability as a function of the total number of communicated bits. We observe that the error exponent  depends on the lattice structure and that zero error requires an infinite number of communicated bits. In contrast, with an infinite number of allowed communication rounds, the nearest lattice point can be determined without  error with a finite average number of communicated bits and a finite average number of rounds of communication. In two dimensions, the hexagonal lattice, which is most efficient for communication and compression,  is found to be the most expensive in terms of communication cost.
\end{abstract}


%
\IEEEpeerreviewmaketitle

\section{Introduction}
\begin{figure}[htbp] 
   \centering
   \includegraphics[width=2.5in]{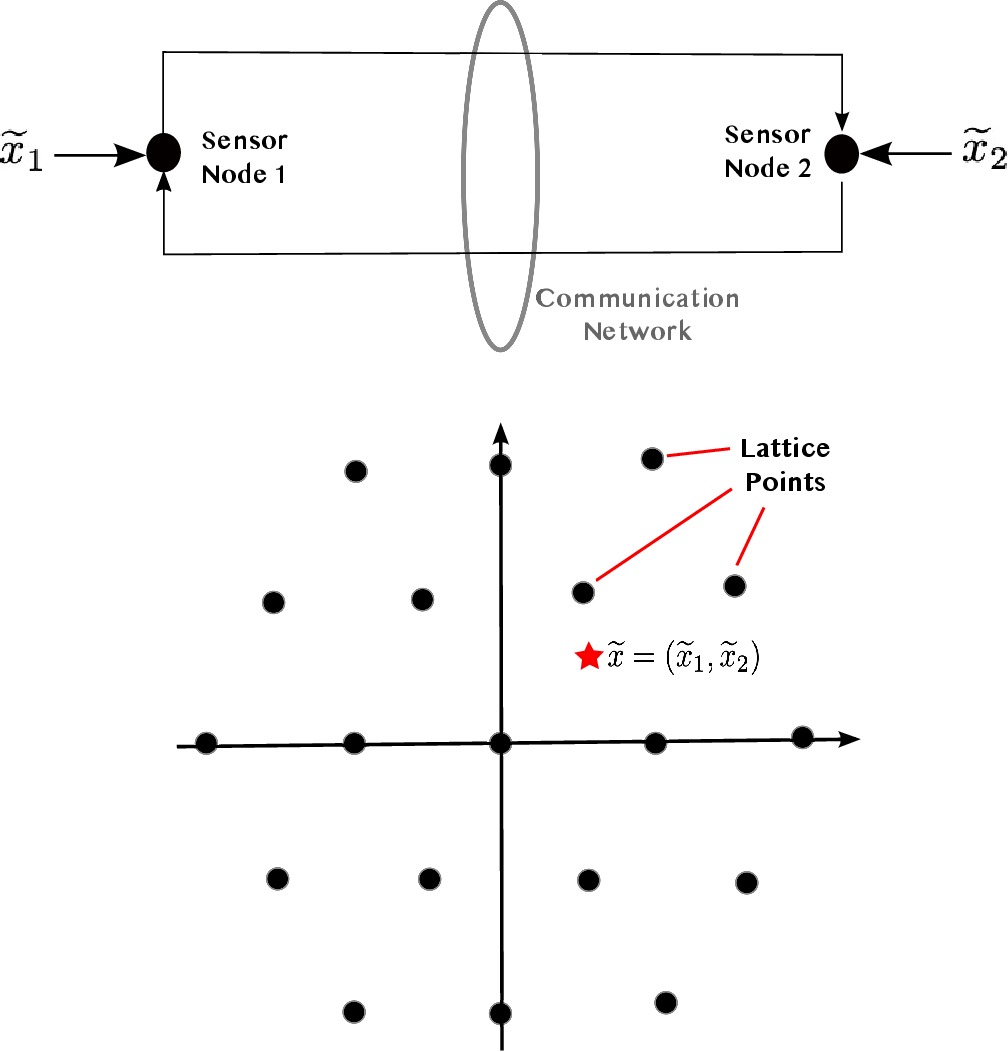}
   \caption{Setup for $n=2$. Two sensors are connected by noiseless links and have partial information about a vector $\tilde{\bm{X}}$. The nodes collaborate in order to find the nearest point in the lattice $\Lambda$. (a) Sensors and communication network (b) signal space visualization in $\mathbb{R}^2$.}
   \label{fig:network1}
\end{figure}

Given a lattice $\Lambda$, which is a discrete additive subgroup of $\mathbb{R}^n$, and a  vector $\bm{x}\in \mathbb{R}^n$, the nearest lattice point problem (NLP) is to determine  the vector $\latpt \in \Lambda$ which is nearest to $\bm{x}$ with respect to the Euclidean norm. NLP is an NP-hard problem~\cite{Boas:1981}. In coding problems, its complexity  limits system performance, while  in cryptography, this  complexity  is harnessed to provide secure communication in the presence of an eavesdropper.

In this paper, we consider a distributed version of the NLP. In this version of the problem there are $n$ sensor nodes connected by a communication network and the $i$th node observes $x_i$, the $i$th component of $\bm{x}$.  The nodes exchange information according to a predefined protocol, such that when the protocol halts each node has computed an approximation to the nearest lattice point for $\bm{x}$.  Our objective is to study the tradeoff between the amount of communication, measured in bits, and the error probability, i.e. the probability that the approximate nearest lattice point does not coincide with the nearest lattice point. Fig.~\ref{fig:network1} illustrates the communication scenario and geometry of the problem for $n=2$.
We restrict our work here to the two-dimensional case, i.e. $n=2$, leaving the study of higher dimensional lattices to a succeeding work. The importance of the two dimensional case and remarks about generalizations are in Sec.~\ref{sec:highdim}.  


Two-party interactive communication has been considered in a series of pioneering papers~\cite{Orlitsky:1990}, \cite{Orlitsky:1991}, \cite{Orlitsky:1992}.  
 Our work is based on analysis techniques for quantization~\cite{Bennet:1948}, \cite{GP:1968}.  Some applications of these quantization methods to distributed detection problems are  in~\cite{Poor:1988},~\cite{Benitz:1989}. 
Distributed classification problems related to the work presented here are in \cite{VVclassifier:2019}, \cite{HEFS:2021}, \cite{DZ2021}. While the problem considered in this paper is of fundamental importance, it also has potential applications to emerging systems for  network security and machine learning, see e.g.~\cite{HEFS:2021},~\cite{li2014communication},\cite{meng2015collaborative} and in distributed MIMO communications~\cite{VV:2021}.

\begin{figure}[htbp] 
   \centering
   \includegraphics[width=2.5in]{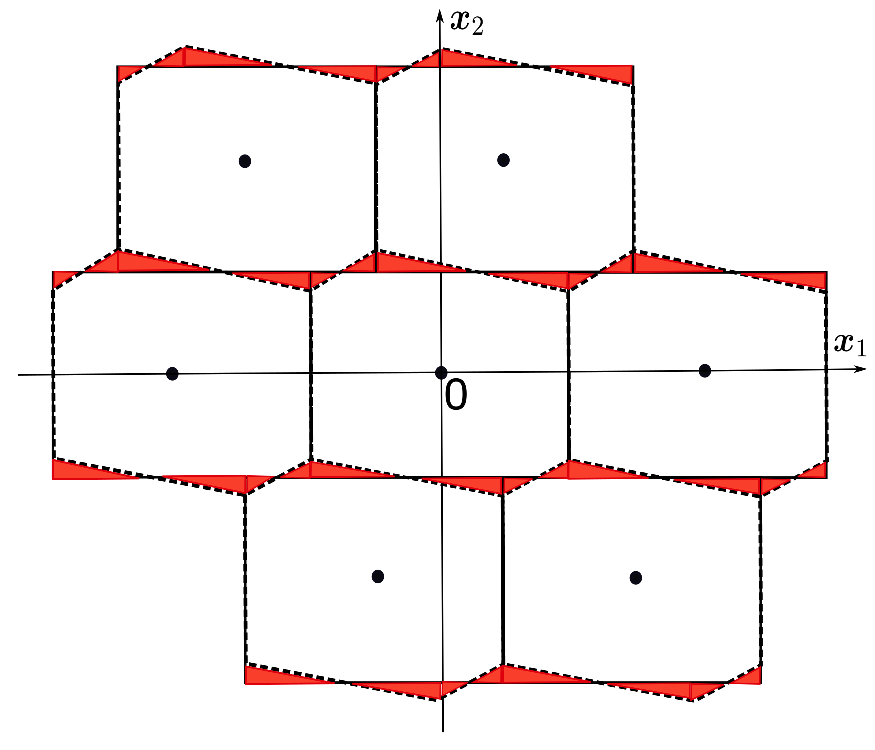} 
   \caption{The Babai partition (rectangular partition with solid lines) and the Voronoi partition (hexagonal partition with dotted lines) for a lattice in $\mathbb{R}^2$. The first stage of the algorithm determines the cell of the Babai partition which contains $\bm{x}$. The second stage reclassifies $\bm{x}$ according to the Voronoi partition. The shaded triangular regions illustrate the points that must be reclassified in the second stage.}
   \label{fig:Bab2Vor1}
\end{figure}
We note that NLP is equivalent to finding an optimum partition (Voronoi partition) of $\mathbb{R}^n$, where each cell of the partition is associated with a unique lattice point and is the set of vectors which are closest to that lattice point (a unique tie-breaking mechanism is assumed). In a companion paper~\cite{BVC:2021}, we have developed a communication protocol and have determined upper bounds for the optimum communication cost of constructing a specific rectangular partition for a given lattice. The partition is referred to as a Babai partition and is an approximation to the Voronoi partition for a given lattice \cite{babai}. Here we develop a two-stage protocol in which the Babai partition computed in the first stage as in~\cite{BVC:2021} is refined in the second stage to yield a Voronoi partition. This is illustrated in Fig.~\ref{fig:Bab2Vor1}, which also shows the points that need to be reclassified.

The main contributions of the paper are as follows. 
\begin{itemize}
\item We consider interactive protocols with single  as well as unbounded number of rounds and study the tradeoff between the error probability (the probability that the protocol fails to recover the nearest lattice point) and the communication cost.
\item For a single round protocol we develop an analytic expression for the tradeoff between the rate and error probability. We show that in the limit as the communication rate grows to infinity, the error probability $P_e$ is given in terms of the communication rate $\Rate_{II}$ (see Remark~\ref{rem:sumrate} in Sec.~\ref{sec:perateoptim}) by
$$
P_e=K_1(\Lambda)e^{-\Rate_{II}/K_2(\Lambda)},
$$
where $K_1$ and $K_2$ are constants that depend on the lattice $\Lambda$ and are specified in Theorem~\ref{thm:optim2}. Specifically $K_1(\Lambda)=\Xi(\theta)$ as defined in \eqref{eqn:optimrwnew}, and $K_2(\Lambda)=1-P_{W_1}(0)$, where $W_1$ is a bin index computed by node 1, as defined in Sec.~\ref{sec:morenotation}, and $P_{W_1}(0)=\prob{W_1=0}$.
\item For interactive protocols with an unbounded number of rounds, we exhibit a construction which results in \emph{zero} error probability with finite average bit cost. This is a surprising result, when compared to the single-round protocol, which can only achieve a strictly positive error probability at a finite rate. 
\item We study the dependence of the communication cost of our protocols on the lattice structure. In particular, we show that the lattice which is best in two dimensions for both coding and quantization, namely, the hexagonal lattice, requires the largest amount of communication in a two-node distributed setting.
\end{itemize}
This work is an extended version of an earlier conference work~\cite{VB:2017}. Compared to this work (i) we are more specific about our source model (ii) provide more details on the structure of the Babai and Voronoi cells, (iii) we describe the partition induced by the communication protocol in more detail, using a general notation, (iv) provide more details of the analysis of the error probability, (v)  provide a proof of optimality for the actions of node 2 (Sec.~\ref{sec:analep}), (vi) optimize the partition and provide a proof of optimality (Sec.~\ref{sec:perateoptim}) and (vii) provide a more detailed numerical exploration of the results of the protocol and the relationship to the geometric structure of the Babai and Voronoi cell (Sec.~\ref{sec:discussion}).

The remainder of the paper is organized as follows.  Mathematical preliminaries and notation are presented in Sec.~\ref{sec:largescale}. The two-dimensional lattices and the source model used in this work is described in Sec.~\ref{sec:sourcemodel}. The interactive single round protocol is  described and notation for the protocol is established   in Sec.~\ref{sec:onedec}. Error probability and communication rate calculations are carried for the single round interactive model in Sec.~\ref{sec:epcommratecalcs} and the partition created by the protocol (equivalently, the quantizer at each node) is optimized in Sec.~\ref{sec:perateoptim}. The interactive protocol with an  unbounded number of rounds of communication is described and performance calculations are carried out in Sec.~\ref{sec:infinitedec}.  Numerical results and a discussion are in Sec.~\ref{sec:discussion}.  Some comments on the significance of the $n=2$ case and generalizations to $n>2$ are in Sec.~\ref{sec:highdim}. A summary and conclusions is provided in Sec.~\ref{sec:summary}.

\section{Mathematical Preliminaries}
\label{sec:largescale}
In this section we fix notation and present definitions  that are relevant to our study.

If ${\mathcal S} \subset \mathbb{R}^n$ and $\bm{x}\in \mathbb{R}^n$, we define ${\mathcal S}+\bm{x}$, the translation of ${\mathcal S}$ by $\bm{x}$,  as the set
$${\mathcal S}+\bm{x}=\{\bm{y}+\bm{x}: \bm{y} \in \mathcal S\}.$$
If $\cB \subset \mathbb{R}^n$, then we define $\cB^{(i)}=\{x_i~:~(x_1,x_2,\ldots,x_n) \in \cB\}$, $i=1,2,\ldots,n$ to be its projection on the $i$th coordinate axis. In case 
$\cB$ is a rectangle, aligned with the coordinate axes, $\cB=\cB^{(1)} \times \cB^{(2)} \ldots \times \cB^{(n)}$.

A collection of subsets of $\mathbb{R}^n$ is called a partition of $\mathbb{R}^n$, and the subsets are referred to as cells of the partition if the union of the cells is $\mathbb{R}^n$ and the interiors of the cells are disjoint.

A (full rank) \emph{lattice} $\Lambda \subset \mathbb{R}^n$ is the set of all integer linear combinations of a set of linearly independent vectors $\{\bm{v}_1,\bm{v}_2,\ldots,\bm{v}_n\} \subset \mathbb{R}^n,$ called a \textit{lattice basis}. Thus
$$\Lambda=\{\bV\bm{u},~\bm{u} \in \mathbb{Z}^n\},
$$ 
where the columns of the \textit{generator matrix} $\bV$ are the basis vectors $\bm{v}_{1}, \dots, \bm{v}_{n}$.

A set $\mathcal{F}$ is called a \textit{fundamental region} of a lattice $\Lambda$ if  its translations by elements of $\latpt \in \Lambda$ cover $\mathbb{R}^{n},$ i.e., $\underset{{\latpt} \in \Lambda}{\bigcup} (\mathcal{F} + {\latpt}) = \mathbb{R}^n,$ and the interiors of ${\latpt}_1 + \mathcal{F}$ and ${\latpt}_2 + \mathcal{F}$ do not intersect for $\latpt_1 \neq {\latpt}_2$. The $n$-dimensional volume of any fundamental region is $\vol\Lambda=|\det \bV|$.  Two kinds of fundamental regions of $\Lambda$ are of interest in this work---the Voronoi cell,  $\Vor$ and a Babai cell $\Bab$, which are described next. 

The \textit{Voronoi region} or \textit{Voronoi cell}   associated with the lattice point $\latpt$, is the set
$$\{\bm{x} \in \mathbb{R}^{n}: \|\bm{x}-\latpt\| \leq \|\bm{x}-{\tilde{\latpt}}\|, \ \text{for all} \ {\tilde{\latpt}} \in \Lambda\},$$ where $\|.\|$ denotes the Euclidean norm. 
The Voronoi cell associated with lattice point $\vzero$ is denoted $\Vor$. The Voronoi cell associated with lattice point $\latpt$ is $  \Vor+  \latpt$.  The  \textit{nearest lattice point problem} consists of finding a lattice point $\latpt_v({\bm x}) \in \Lambda$ that minimizes the Euclidean distance to ${\bm x} \in \mathbb{R}^n,$ i.e., 
\[
\latpt_v({\bm x}) = \arg \min_{\latpt \in \Lambda} \| \bm{x}-\latpt \|^{2}.
\]
The solution to  the nearest lattice point problem partitions $\mathbb{R}^n$ into Voronoi cells, as  described above, and the resulting partition of $\mathbb{R}^n$ is called its Voronoi partition.

Given a lattice with an ordered basis $\{\bv_1,\bv_2,\ldots,\bv_n\}$ the associated Babai cell is a rectangular axis-aligned fundamental region determined by the basis. The Babai cell is  easiest to describe if we assume that the generator matrix is  upper triangular. Note that  a triangular generator matrix can always be obtained after applying the QR decomposition to the original generator matrix. Given a lattice $\Lambda \subset \mathbb{R}^n$ with an upper triangular generator matrix $\bm{V}$,  and $\bm{x}=(x_1,x_2,\ldots,x_n)$, the Babai point is obtained  as follows. First $\bu=(u_1,u_2,\ldots,u_n)$ is computed according to
	\begin{equation}\label{eqn:BabaiU}
{u}_i=\mynint{\frac{x_i-\sum_{j=i+1}^nv_{i,j}{u}_j}{v_{i,i}}}
	\end{equation}
in the order $i=n,n-1,\ldots,1$ and where $\mynint{x}$ denotes the nearest integer to $x$. The Babai point is then given by
 \begin{equation}
 \latpt_{np}({\bm x})= \bV \bu.
 \label{eqn:BabaiPoint}
 \end{equation}
The set of ${\bm x} \in \mathbb{R}^n$ mapped to $\latpt_{np} \in \Lambda$ by (\ref{eqn:BabaiU}) and (\ref{eqn:BabaiPoint}) is called the Babai cell associated with lattice point $\latpt_{np}$. The Babai cell corresponding to the lattice point $\vzero$ is denoted $\Bab$, which is a rectangle with vertices $(\pm v_{1,1}/2,\pm v_{2,2}/2,\ldots,\pm v_{n,n}/2)$.  The Babai cell associated with lattice point $\latpt$ is $\Bab+\latpt$. The Babai cells partition $\mathbb{R}^n$, and the partition is referred to as the Babai partition. Note that since  $\Bab$ is rectangular and axis-parallel, $\Bab= \Bab^{(1)}\times \Bab^{(2)}\times\ldots\times \Bab^{(n)}$. For a general two dimensional lattice, the Voronoi and  Babai partitions are illustrated in Fig.~\ref{fig:detailedGeom}.

Since the $\Bab$ and $\Vor$ are both fundamental regions of lattice $\Lambda$, it is true that for any $\latpt',\latpt'' \in \Lambda$, $(\Bab+{\latpt'})\bigcap(\Vor+{\latpt''})=(\Bab\bigcap(\Vor+{\latpt''-\latpt'})) +\latpt'$, i.e. the joint Babai and Voronoi partition is invariant to lattice translations. This translation invariance is exploited by our algorithms in order to reduce the complexity.

The length of an interval $\mathcal{I}\subset \mathbb{R}$ is denoted $|\mathcal{I}|$.
For a differentiable function $f~:~\mathbb{R} \to\mathbb{R}$, $\dot{f}(x)$ will denote its  slope at $x$. 
Random variables and vectors and their realizations, are denoted by upper and lower case versions of the same letter, respectively.  The entropy function is denoted by $\ent{X}$, where $X$ is a random variable, or by $H(P)$, where $P$ is a probability distribution. Vectors and matrices are  written in boldface.

\section{Lattice Basis and Source Model}
\label{sec:sourcemodel}

We now describe the family of lattices  and the statistical model for the information source used in this work. We will consider the two-node case, and lattices of dimension two. 
\subsection{Lattice Basis}
The  lattice $\Lambda$  is assumed to have an upper triangular generator matrix $\bV$ of  the  form 
 \begin{equation}
 \bV=\begin{pmatrix} 1 & \rho \cos \theta \\ 0 & \rho \sin \theta\end{pmatrix}.
 \label{eqn:GenMatrix}
 \end{equation}
 The restriction on the parameters $\rho$ and $\theta$ is explained next.

\begin{figure}[htbp] 
   \centering
 \includegraphics[width=2.5in]{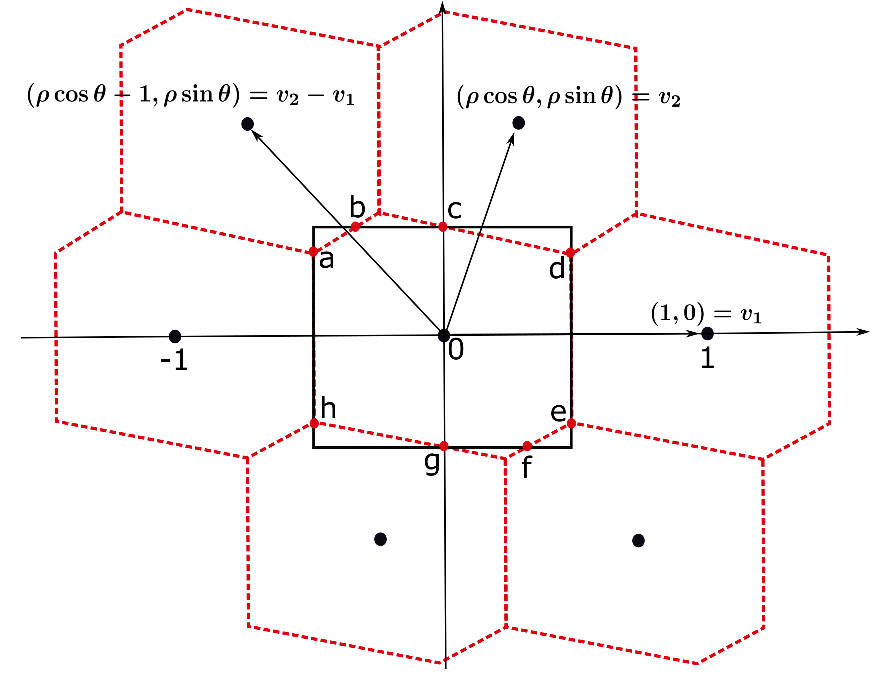}  
   \caption{$\Bab$ and intersecting Voronoi cells for a given lattice. Coordinates of lattice basis vectors $\bv_1$, $\bv_2$ and face intersection points  $\bm{a},\bm{b},\ldots,\bm{h}$, of  Babai and Voronoi cells are shown. Faces of the Voronoi cell $\Vor$ are determined by the relevant vectors, $\pm \bv_1$, $\pm \bv_2$ and $\pm (\bv_2-\bv_1)$. Coordinates of cell boundary intersection points: $\bm{b}=(-1+\rho \cos \theta,\rho \sin \theta)/2$, $\bm{c}=(\rho \cos \theta,\rho \sin \theta)/2$, $\bm{d}=(1,(\rho-\cos \theta)/\sin \theta)/2$, $\bm{e}=(1,-(\rho-\cos \theta)/\sin \theta)/2$, $\bm{f}=-\bm{b}$, $\bm{g}=-\bm{c}$, $\bm{h}=-\bm{d}$ and $\bm{a}=-\bm{e}$.}
   \label{fig:detailedGeom}
\end{figure}

A basis $\{\bm{v_{1}},\bm{v_{2}}\}$ of a lattice $\Lambda \subset \mathbb{R}^{2}$ is said to be \textit{Gauss-Lagrange reduced}\footnote{A generalization of Gauss-Lagrange reduced basis to higher dimensions leads to the definition of a \textit{Minkowski-reduced} basis.}
if $\bv_1$ is the shortest (non-zero) vector in the lattice and $\bv_2$ is the shortest vector for which $\{\bv_1,\bv_2\}$ is a basis for $\Lambda$. The basis given by the columns of \eqref{eqn:GenMatrix} is Gauss-Lagrange reduced if and only if  $-1< 2\rho \cos \theta \leq 1 \leq \rho^2$~\cite{SPLAG}. Further, the \textit{relevant vectors}, i.e. the lattice vectors  that define the faces of the Voronoi cell $\Vor$ are  $\pm(1,0), \pm(\rho \cos \theta, \rho \sin \theta)$ and $\pm (\rho \cos \theta -1,\rho \sin \theta)$   and it suffices to restrict $ \tfrac{\pi}{3} \leq \theta \leq \tfrac{\pi}{2}$. 
More details are in Fig.~\ref{fig:detailedGeom}. We will consider lattices with these restrictions in this work.

The Babai point $\latpt_{np}({\bm x})=\bV\bu$, $\bu=(u_1,u_2)$, obtained from \eqref{eqn:BabaiU}, is   given more explicitly by
\begin{equation}\label{eqn:Babai2D}
u_2 = \mynint{\frac{x_2}{\rho \sin\theta}}, ~ u_1= \mynint{x_1 - (\rho \cos\theta) u_2 }.
\end{equation}

\subsection{Source Model}
Suppose we are given a source $\tilde{\tilde{\bm{X}}}=(\tilde{\tilde{X}}_1,\tilde{\tilde{X}}_2)$, with source pdf $p_{\tilde{\tilde{\bm{X}}}}$. For a given lattice $\Lambda$ with basis $\bV$, we will work with a derived source, whose pdf is  constant over every cell of the  Babai partition of $\mathbb{R}^2$. Specifically we will work with a source 
 vector  $\tilde{\bm{X}}=(\tilde{X}_1,\tilde{X}_2)$   whose pdf  
 \begin{equation}
 p_{\tilde{\bm{X}}}(\bm{x})=p_\latpt,~{\bm{x}}\in \Bab+\latpt,~\latpt \in \Lambda.
 \end{equation}
 where
 \begin{equation}
 p_\latpt=\frac{\prob{\tilde{\tilde{\bm{X}}}\in \Bab+\latpt}}{\vol{\Bab}},~ \latpt \in \Lambda.
 \end{equation}


\section{Description of the Interactive Single Round Protocol}
\label{sec:onedec}

\begin{figure}[htbp] 
   \centering   
   \includegraphics[width=3in]{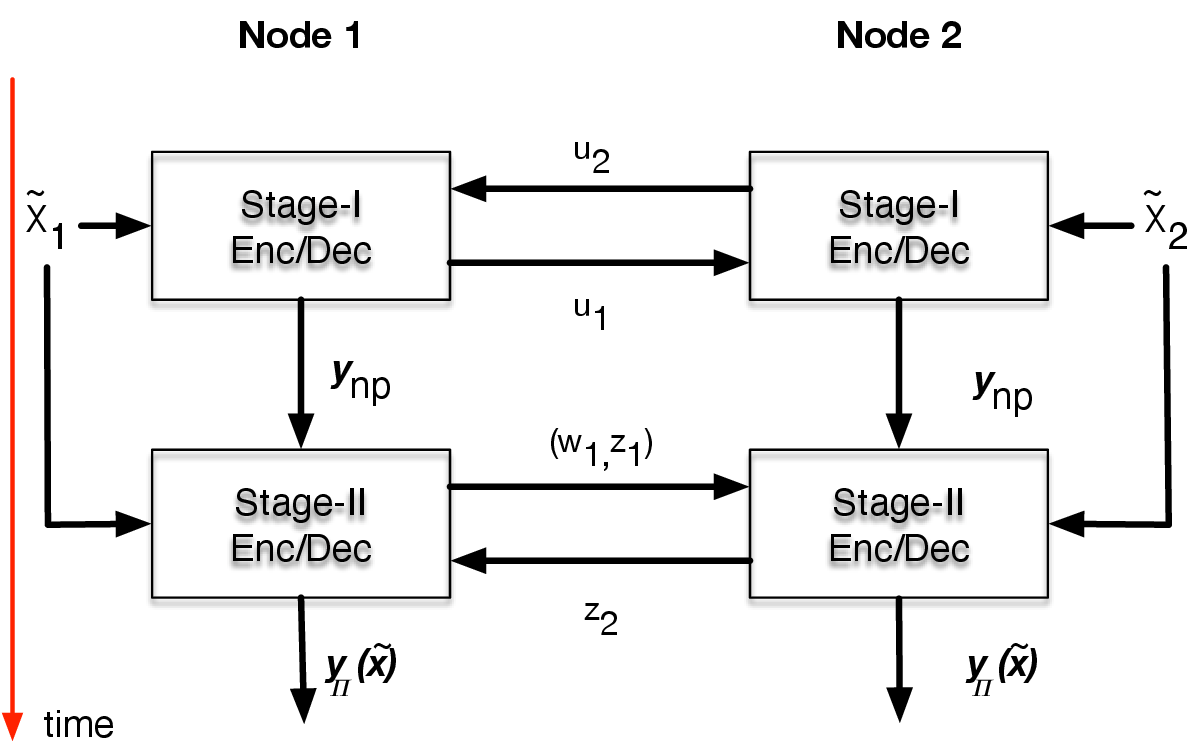} 
   \caption{Operational description of the two-stage protocol for finding an approximation to  $\latpt_{v}(\tilde{\bm x})$. At the end of Stage-I, both nodes have determined $\latpt_{np}(\tilde{\bm x})$. Stage-II then refines the result of Stage-I, resulting in the lattice point $\bm{y}_\Pi(\tilde{\bm{x}})$.  }
   \label{fig:twostage}
\end{figure}
In a two-node network,  node $i$ observes random variable $\tilde{X}_i$, $i=1,2$. As mentioned earlier, the objective is for the nodes to exchange information in an interactive manner, according to a pre-established protocol $\Pi$,  such that both nodes possess  an approximation to $\latpt_{v}(\bm{\tilde{\bm{x}}})$. This approximation is denoted  ${\latpt}_\Pi(\tilde{\bm{x}})$, and is the lattice point obtained when communication stops. Our protocol operates in two stages, Stage-I and Stage-II. An operational  illustration is in Fig.~\ref{fig:twostage}. The separation of the protocol into two stages is in order to exploit the translation invariance of the lattice with respect to the lattice basis vectors and uses the fact that the  Babai and Voronoi cells are both fundamental regions for the lattice. The first stage determines the Babai point ${\latpt}_{np}(\tilde{\bm{x}})$ associated with the vector $\tilde{\bm{x}}=(\tilde{x}_1,\tilde{x}_2)$. The second stage computes $\latpt_\Pi(\tilde{\bm{x}})$.

Communication in Stage-I  proceeds as follows. Node-2 observes $\tilde{x}_2$ and sends integer $u_2$ to node-1, where $u_2$ is given by \eqref{eqn:Babai2D}. Node-1, based on observing $\tilde{x}_1$ and $u_2$, computes $u_1$ according to \eqref{eqn:Babai2D}, and sends this to node-2. At the conclusion of this stage of the protocol, both nodes have determined  ${\latpt}_{np}(\tilde{\bm{x}})$, thus localizing $\tilde{\bm{x}}$ to the Babai cell $\Bab+{\latpt_{np}}$.

The Babai point ${\latpt}_{np}(\tilde{\bm{x}})$ is then subtracted off in Stage-II, which first computes $\bx=\tilde{\bx}-\latpt_{np}(\tilde{\bx})$. The Stage-II encoder then proceeds to encode $\bx$, using a quantizer which is independent of the Babai point. This  significantly reduces the complexity of the encoding and decoding algorithms. Stage-II determines the lattice point $\latpt_\Pi(\bx)$ and ${\latpt}_\Pi(\tilde\bx)=\latpt_\Pi(\bx)+\latpt_{np}(\tilde\bx)$.

\begin{figure}[htbp] 
   \centering
   \includegraphics[width=3.0in]{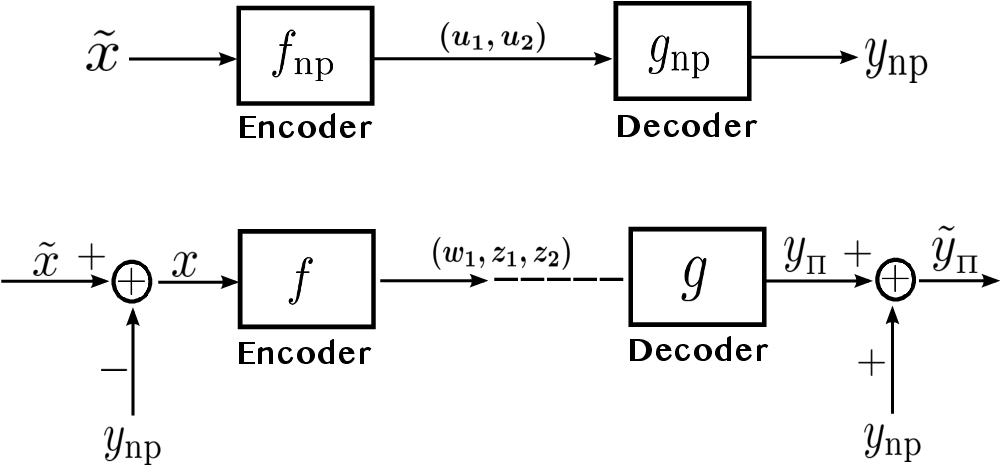} 
   \caption{The collective action of  encoding and decoding operations in Stage-I (top) and Stage-II (bottom). In Stage-II the vector  $\bm{x}$ always lies in the Babai cell $\Bab$.  }
   \label{fig:transinv}
\end{figure}

An illustration of both stages of the protocol,  which  shows  explicit encoders and decoders for Stage-I and Stage-II,  is in Fig.~\ref{fig:transinv}. 
In this figure, the collective action (i.e. of  both nodes) of the encoders (decoders)  in Stages-I and -II is denoted $f_{np}$ and $f$  ($g_{np}$ and $g$), respectively.

We have already described $f_{np}$ and $g_{np}$. We now describe $f$ and $g$ of Stage-II of the protocol $\Pi$ in more detail. For this, we first describe the structure of $\Bab$ and the Voronoi cells that intersect it in Sec.~\ref{sec:babvorgeom}, followed by a detailed description of the encoder $f$ in Sec.~\ref{sec:morenotation} and the decoder $g$ in Sec.~\ref{sec:decoder}.

\subsection{Geometric Structure of $\Bab\bigcap \Vor$}
\label{sec:babvorgeom}

The input to the Stage-II encoder $f$ is  $\bm{x}=(x_1,x_2)$, where $x_1$ is the horizontal coordinate and $x_2$ is the vertical coordinate. By construction, $\bm{x}$ is uniformly distributed over the Babai cell $\Bab$. Since some $\bm{x}$ will lie in Voronoi cells associated with non-zero lattice points, it is useful to understand the geometric structure of the Babai cell $\Bab$ and the Voronoi cells $\Vor+\latpt,~\latpt\in \Lambda$ which intersect it.

The six face-determining lattice points for $\Vor$ are $\pm \bv_1$, $\pm \bv_2$ and $\pm \bv_3$, where  $\bv_1=(1,0)$, $\bv_2=(\rho \cos \theta, \rho \sin \theta)$ and $\bv_3=(-1+\rho \cos \theta,\rho \sin \theta)$, as seen in  Fig.~\ref{fig:detailedGeom}. The walls of the Voronoi cell $\Vor$ are $\bm{x}^t \bv_i=\|\bv_i\|^2/2$, $\bm{x}^t \bv_i=-\|\bv_i\|^2/2$, $i=1,2,3$. The walls of the Babai cell $\Bab$ are described by the equations $x_1=1/2$, $x_1=-1/2$, $x_2=\rho \sin \theta/2$ and $x_2=-\rho \sin \theta /2$. Thus $|\Bab^{(1)}|=\lengthbabai=1$ and $|\Bab^{(2)}|=\heightbabai=\rho \sin \theta$.
The boundary of the intersection, $\Bab\bigcap \Vor$ plays an important role in the analysis. Let $x_2=u(x_1)$ describe its upper boundary and $x_2=l(x_1)$ describe its lower boundary, where $x_1\in \Bab^{(1)}$. For all lattices with generator matrix \eqref{eqn:GenMatrix}, $u(x_1)$ and $l(x_1)$ are piecewise linear in $x_1$ and are determined by the boundaries of the Babai and Voronoi cells. 
We can express the functions, respectively,  as follows:
\begin{IEEEeqnarray}{lCr}\label{eq:u}
u(x_1) = \begin{cases}
 \left(\frac{-\rho\cos\theta +1}{\rho\sin\theta} \right)x_1 + \left(\frac{\rho^2-2\rho \cos \theta+1}{2\rho \sin \theta} \right), ~ x_1 \in \left[\tfrac{-1}{2}, \tfrac{\rho\cos\theta-1}{2} \right), \nonumber \\
\frac{\rho\sin\theta}{2},  ~ x_1 \in  \left[\tfrac{\rho\cos\theta-1}{2},  \tfrac{\rho\cos\theta}{2}\right), \nonumber \\
\left(-\frac{\cos\theta}{\sin\theta}\right)x_1  + \left(\frac{\rho}{2\sin \theta} \right),    ~x_1 \in \left[ \tfrac{\rho\cos\theta}{2} ,\tfrac{1}{2} \right),
\end{cases}
\end{IEEEeqnarray}
and
\begin{IEEEeqnarray}{lCr}\label{eq:l}
l(x_1) = \begin{cases}
\left(-\frac{\cos\theta}{\sin\theta}\right)x_1 - \frac{\rho}{2\sin \theta} , ~ x_1 \in \left[ -\tfrac{1}{2},- \tfrac{\rho\cos\theta}{2} \right),  \nonumber \\
-\frac{\rho\sin\theta}{2},    ~x_1 \in \left[- \tfrac{\rho\cos\theta}{2} , \tfrac{-\rho\cos\theta+1}{2} \right), \nonumber \\
\left(\frac{-\rho\cos\theta +1}{\rho\sin\theta} \right)x_1 - \left( \frac{\rho^2-2\rho \cos \theta +1}{2\rho \sin \theta}\right),~  x_1 \in \left[ \tfrac{-\rho\cos\theta+1}{2} , \tfrac{1}{2} \right).
\end{cases}
\end{IEEEeqnarray}
Recall that  $0 < \rho \cos \theta < 1/2.$ Details of $\Bab\bigcap \Vor$ are in  Fig.~\ref{fig:detailedGeom}.

From the above description we observe that four  walls of $\Bab\bigcap\Vor$ are  determined by the walls of the Babai cell, a pair each of vertical and horizontal walls.  Of these, the vertical walls grow in size, while the horizontal walls shrink, as $\theta$ increases. Four other walls of $\Bab \bigcap \Vor$ are determined by $\Vor$. This can be seen in Fig.~\ref{fig:detailedGeom}. 

As will become clear in the Sec.~\ref{sec:analep}, the quantity $|\dot{u}(x_1)|+|\dot{l}(x_1)|$ plays an important role in the performance analysis\footnote{Recall that $\dot{u}$  is the derivative of a real function $u$.}. We observe here that $|\dot{u}(x_1)|+|\dot{l}(x_1)|$ only takes on  a finite number of distinct values as $x_1$ varies over $\Bab^{(1)}$.

%
%
%

A detailed description of the Stage-II protocol and the partition of $\Bab$ that it creates is now presented.

\subsection{Stage-II Protocol Description: Further Details}
\label{sec:morenotation}
We describe the encoder  $f$, which maps $\bm{x}=(x_1,x_2)\in \Bab$ to $(w_1,z_1,z_2)$, $w_1\in \{0,1,\ldots,m\}$, $z_1\in \{0,1,\ldots,N_{w_1}-1\}$, $z_2\in \{-k_{w_1,z_1},\ldots,k_{w_1,z_1}\}$ for a given set of non-negative integers $m$, $\{N_{w_1}, w_1=0,1,\ldots,m\}$, and $\{k_{w_1,z_1},~w_1=0,1,\ldots,m,~z_1=0,1,\ldots,N_{w_1}-1\}$. In more detail, the range of $x_1$ is partitioned into $m$ bins; the bin index is given by $w_1$. The bin identified by $w_1$ is further partitioned into $N_{w_1}$ bins, and finally, given $w_1$ and $z_1$, the range of $x_2$ is partitioned into $2k_{w_1,z_1}+1$ bins. 

\begin{figure}[htbp] 
   \centering
   \includegraphics[width=3.0in]{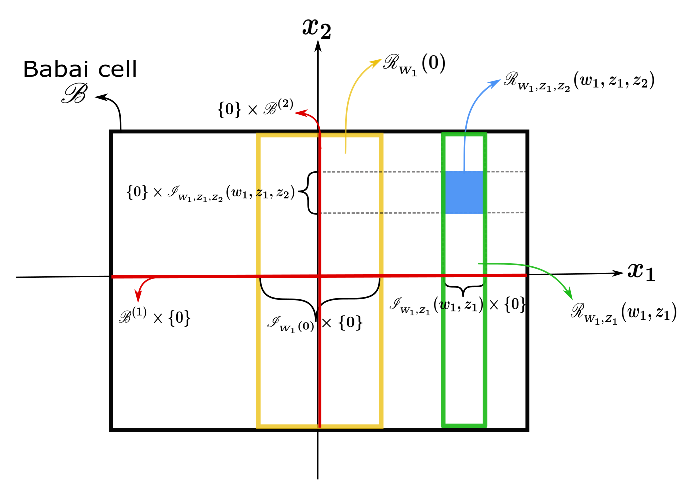} 
   \caption{Visual depiction of the notation developed for describing intervals and rectangles in the partition of $\Bab$ by nodes 1 and 2 in Stage-II of the protocol.}
   \label{fig:notationdetails1}
\end{figure}
\begin{figure}[htbp] 
   \centering
   \includegraphics[width=3.0in]{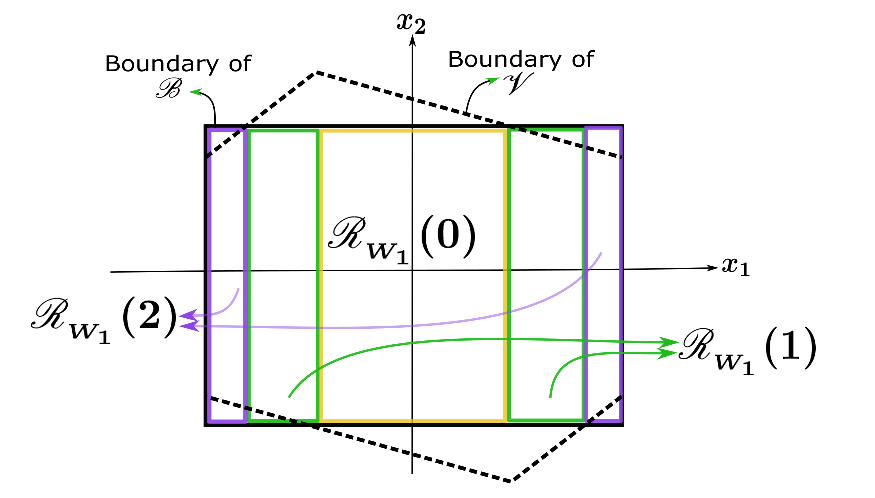} 
   \caption{Partition of $\Bab$ created by $W_1$. Note that $\Rone$  is a union of disjoint rectangles for $w_1=1$ and $w_1=2$.}
   \label{fig:notationdetails2}
\end{figure}

Some additional notation is needed to describe the bins created by nodes 1 and 2. The reader will find it helpful to consult Figs. ~\ref{fig:notationdetails1} and \ref{fig:notationdetails2}. Node-1 first computes $w_1$ based on $x_1$, effectively partitioning $\Bab^{(1)}$ into  regions $\Intone,~w_1=0,1,\ldots,m$, such that  for any $x_1\in \Intone$,  $|\dot{u}(x_1)|+|\dot{l}(x_1)|$ is constant and $m$ is the smallest positive integer for which such a partition into regions of constant sum absolute slope can be obtained.  $\Intonezero$ is special, because for $x_1\in \Intonezero$, the line segments  $x_2=u(x_1)$ and $x_2=l(x_1)$ coincide with the boundary of $\Bab$ and thus $|\dot{u}(x_1)|+|\dot{l}(x_1)|=0$.   Each region  $\Intone,~w_1\neq 0$, is subdivided further into bins, specifically $\Intone$ is divided  into $N_{w_1}$ bins, denoted $\Inttwo$,  where  $z_1\in \{0,1,\ldots,N_{w_1}-1\}$. $W_1$ and if necessary, $Z_1$, are sent to node 2.
 At this point, i.e. after node-1 has concluded its transmission, $\Bab^{(1)}$, has been  partitioned into bins, $\Intonezero$ and  $\Inttwo$, where $w_1 \in \{1,2,\ldots,m\}$  and $z_1\in \{0,1,\ldots,N_{w_1}-1\}$ and thus  the Babai cell $\Bab$ has been partitioned into rectangles 
 \begin{IEEEeqnarray*}{lCl}
\{\Ronezero=\Intonezero\times \Bab^{(2)},  \\
 ~~\Rtwo=\Inttwo\times \Bab^{(2)},  \\
 ~~~ w_1=1,2,\ldots,m,~z_1=0,1,\ldots,N_{w_1}-1\}.
 \end{IEEEeqnarray*}
 Fig.~\ref{fig:notationdetails1} illustrates the notation used for describing the partition of $\Bab$ created by nodes 1 and 2.  
 
Now  consider the actions of node 2. If $w_1=0$, then no further communication is necessary and both nodes  compute $\latpt_{\Pi}(\bx)=\bm{0}$, since $\bx\in\Bab\bigcap\Vor$ for all possible values of $x_2$.  Now suppose $W_1=w_1\neq 0$ and $Z_1=z_1$. Node 2 partitions $\Bab^{(2)}$ into $2k_{w_1,z_1}+1$ intervals denoted $\Intthree$, $z_2\in\{-k_{w_1,z_1},\ldots,k_{w_1,z_1}\}$. We define the rectangle $\Rthree=\Inttwo\times\Intthree$. 

We denote by $\lenone$, $\lentwo$ and $\lenthree$  the  measure (the length if an interval)  of the sets $\Intone$, $\Inttwo$ and $\Intthree$, respectively.

Finally, we emphasize that the event $$\{W_1=w_1,Z_1=z_1,Z_2=z_2,\Latpt_{np}=\latpt\}$$ is equivalent to $$\bX\in\Rthree+\latpt.$$


\subsection{Decoder}
\label{sec:decoder}
We now describe the decoder  $g$ which maps $(w_1,z_1,z_2)$ to $\Lambda$. Note that both nodes know $(w_1,z_1,z_2)$ when communication stops and can implement $g$.
We will assume that $g$ is  a \emph{majority} decoder, which sets 
$$
g(w_1,z_1,z_2)=\latpt',~\latpt'\in \Lambda,
$$
if
\begin{IEEEeqnarray*}{rCl}
 \vol{\Rthree\bigcap (\Vor+\latpt')} \geq  \nonumber \\
 \vol{\Rthree\bigcap (\Vor+\latpt'')}
\end{IEEEeqnarray*}
for all $\latpt'' \in \Lambda$. Here we use $\vol{}$ in a generalized sense. In two dimensions, this is the area measure of a set.
Note that $\latpt_\Pi(\bm{x})=g(f(\bm{x}))$, where $f$ and $g$ are as in Fig.~\ref{fig:transinv}.

\section{Error Probability and Communication Rate Calculations}
\label{sec:epcommratecalcs}
We will derive a general expression for the error probability in Sec.~\ref{sec:ep}. This allows us to optimize the actions of node-2 in Sec.~\ref{sec:analep} after which a partially optimized expression for the error probability is obtained in Sec.~\ref{sec:paropep}. General expressions for the communication rates are obtained in Sec.~\ref{sec:commrates}. This sets the stage for optimization of the partition subject to constraints on the communication rate in Sec.~\ref{sec:perateoptim}. The expressions derived here can be applied to the order\footnote{Order $ij$ means node-$i$ transmits first, followed by node-$j$.} $12$ and $21$, except that the coordinates $x_1$ and $x_2$ must be interchanged when the order $21$ is performed.

\subsection{Error Probability}
\label{sec:ep}
Let
\begin{equation*}
\tilde{\cE}=\{\tilde\bx\in \mathbb{R}^2 \mid \latpt_\Pi(\tilde\bx-\latpt_{np}(\tilde\bx))+\latpt_{np}(\tilde\bx)\neq \latpt_v(\tilde\bx)\}.
\end{equation*}
$\prob{\tilde\bX\in \tilde{\cE}}$ is the error probability of our protocol. We show that this error probability, which depends on the infinitely large error region $\tilde{\cE}$, can be reduced to an error probability calculation over the single Babai cell $\Bab$.
Towards this end, let 
\begin{equation*}
{\cE}=\{\bm{x}\in \Bab \mid \latpt_\Pi(\bm{x})\neq \latpt_v(\bm{x})\}.
\end{equation*}
Then
\begin{eqnarray}
P_e & = & \prob{\tilde{\bm{X}} \in \tilde{\cE}} \nonumber \\
& =& \prob{\Latpt_\Pi(\bX)\neq \Latpt_v(\tilde\bX)-\Latpt_{np}(\tilde\bX)} \nonumber \\
&  \stackrel{(a)}{=} & \prob{\Latpt_\Pi(\bX)\neq \Latpt_v(\bm{X})} \nonumber \\
& = & \prob{\bm{X} \in \cE}
\end{eqnarray}
where in (a) we use the fact that the Voronoi cell is a fundamental region of the lattice.
Based on the detailed description of the protocol, we can express the error probability by
\begin{IEEEeqnarray}{lCl}
\prob{\bm{X}\in \cE}  =  \nonumber \\
\sum_{w_1=0}^m \sum_{z_1=0}^{N_{w_1}-1}\sum_{z_2=-k}^k\prob{\bm{X} \in \cE\bigcap \Rthree}.~\label{eqn:bigprob}
\end{IEEEeqnarray}
We now optimize the actions of node-2 to minimize the inner summation  in \eqref{eqn:bigprob}, namely,
$$\sum_{z_2=-k}^k\prob{\bm{X} \in \cE\bigcap \Rthree}.$$

\subsection{Optimizing the Action of Node-2}
\label{sec:analep}


%
%
\begin{figure}[htbp] 
   \centering
   \includegraphics[width=2.5in]{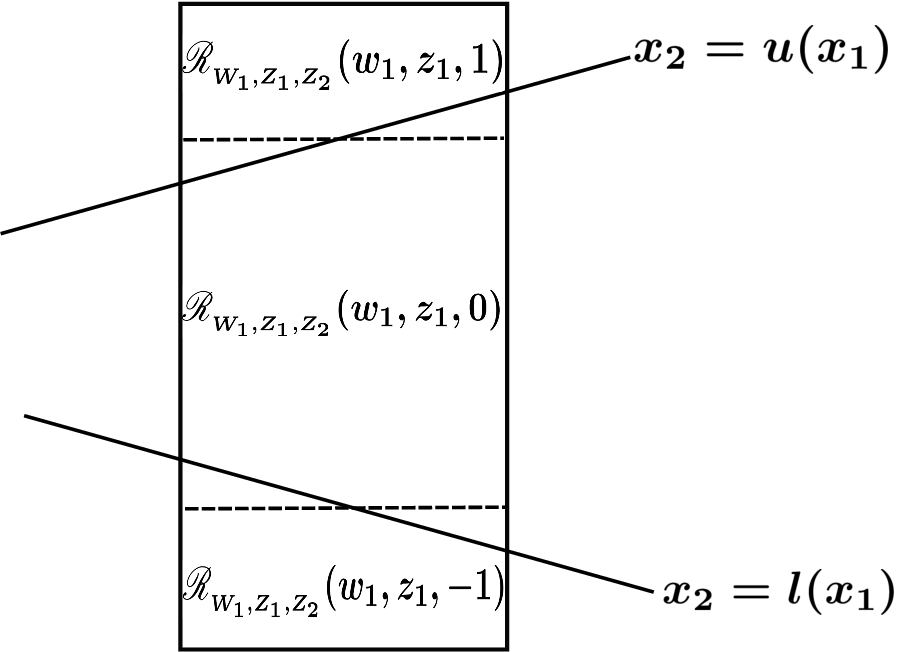} 
   \caption{A  typical partition of  $\Rtwo$  into three sub-rectangles  $\Rthree,~z_2\in \{-1,0,1\},~w_1\neq 0$.}
\label{fig:verticalslice}
\end{figure}

We have already argued that if $W_1=0$, no further communication is required from node-2, and $\latpt_\Pi(\bx)=\latpt_v(\bx)=0$ and the error probability is zero. Our objective is to quantize $X_2$, conditioned on $W_1=w_1\neq0$ and $Z_1=z_1$, or equivalently to partition the interval $$\Bab^{(2)}=[-|v_{2,2}|/2,|v_{2,2}|/2]$$ into intervals $\Intthree, z_2=-k_{w_1,z_1},\ldots,k_{w_1,z_1}$. The reader is referred to Fig.~\ref{fig:verticalslice}. We now show that it is sufficient to set $k_{w_1,z_1}=1$.

In the theorem that follows it  is important to recall that we are assuming a majority decoder, and that the decoder estimate of the Voronoi point is optimized for each of the rectangles $\Rthree=\Inttwo\times \Intthree$, $z_2=-k_{w_1,z_1},\ldots,k_{w_1,z_1}$.
\begin{theorem}\label{thm:optimalnode2}
For any partition of $\Bab^{(2)}$ into intervals $\Intthree,~z_2=-k_{w_1,z_1},\ldots,k_{w_1,z_1}$, for given $W_1=w_1\neq 0$ and $Z_1=z_1$
\begin{IEEEeqnarray}{rCl}
\sum_{z_2=-k_{w_1,z_1}}^{k_{w_1,z_1}}\prob{ \bm{X} \in \cE \bigcap \Rthree}   \nonumber \\  \geq  \frac{(|\dot{u}|+|\dot{l}|)(\lentwo)^2}{4\lengthbabai \heightbabai} ~ ~ ~ ~ ~ \label{eqn:prob}
\end{IEEEeqnarray}
where $\dot{u}$ and $\dot{l}$ are the  (constant) slopes of $u(x_1)$ and $l(x_1)$ for $x_1\in \Inttwo$. 
Further, equality holds when $k_{w_1,z_1}=1$ and the two horizontal cuts are located at $x_2=u(x_1^*)$ and $x_2=l(x_1^*)$, where $x_1^*$ is the midpoint of the interval $\Inttwo$. 
\end{theorem}
\begin{proof}
The reader is referred to Fig.~\ref{fig:AD}.  To simplify the description we refer to $\Rtwo$ as $\Rblank$, and its  length $\lentwo$ as $\delta$ and to $k_{w_1,z_1}$ by $k$. $\Rblank$ intersects the upper boundary $x_2=u(x_1)$  as shown. For simplicity we argue based on the upper boundary $x_2=u(x_1)$. The argument for the lower boundary $x_2=l(x_1)$ is similar. Hereafter, we refer to the upper boundary $x_2=u(x_1)$ as the line. 

Fig.~\ref{fig:AD} shows   a sequence of $2k$ horizontal cuts made by node-2, located at $x_2=t_{-k},~x_2=t_{-k+1},\ldots,x_2=t_{k-1}$.  These cuts partition $\Rblank$ into $2k+1$ sub-rectangles, $\Rthree$ and $z_2=j$ if $j$ is the smallest for which $x_2 \leq t_j$ and $z_2=k$ if $x_2>t_{k-1}$. The only sub-rectangles that contribute to the error probability are the ones which are intersected by the line. Recall the majority decoding rule, which assigns to each sub-rectangle the lattice point whose Voronoi cell has maximal intersection area with that sub-rectangle. The volume contribution from each such rectangle is from one the parts of the sub-rectangle separated by the line, the left part, or the right part. It follows, from the orientation of the line, that if the left part of the rectangle is an error region, the left parts of all rectangles below it are error regions, and if the right part of a rectangle is an error region, then, the right part of every rectangle above it is an error region. Some allowed and disallowed configurations are shown in Fig~\ref{fig:AD}. 

\begin{figure}[htbp] 
   \centering
   \includegraphics[width=2.5in]{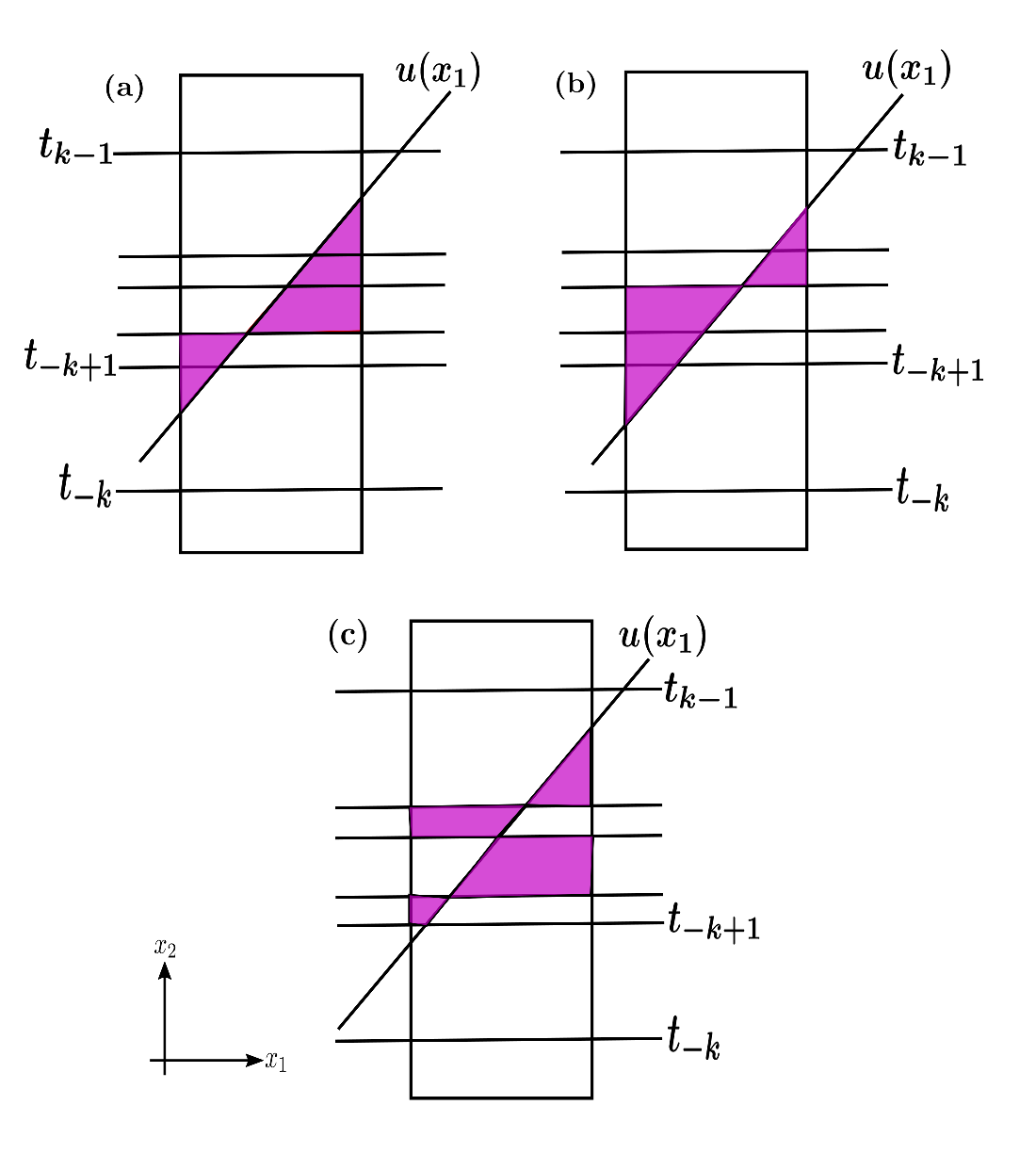} 
   \caption{Allowed and disallowed configurations that arise in the proof of Theorem~\ref{thm:optimalnode2}. Error causing triangles and paralellopipeds are shaded. Configurations  (a) and (b) are allowed , while (c) is disallowed.}
   \label{fig:AD}
\end{figure}

The union of the error regions from all the sub-rectangles reduces to two triangles, as illustrated in the allowed configurations in Fig.~\ref{fig:AD}.
The length of the base of each of the two triangles is $\alpha\delta$, $(1-\alpha)\delta$,  where $0\leq \alpha  \leq 1$, and the corresponding areas of the triangles are $(1/2)\alpha^2\delta^2|\dot{u}|$, $(1/2)(1-\alpha)^2\delta^2|\dot{u}|$.   Since $\alpha^2+(1-\alpha)^2$ is minimized at $\alpha=1/2$ and the minimum value is $1/2$ it follows immediately, by applying a similar argument to the lower boundary $x_2=l(x_1)$, that,
\begin{IEEEeqnarray}{rCl}
\sum_{z_2=-k}^k \prob{ {\bm X} \in \cE \bigcap \Rthree}  \nonumber \\  \geq \frac{\delta^2(|\dot{u}|+|\dot{l}|)}{4\lengthbabai \heightbabai}, ~ ~ ~
\end{IEEEeqnarray}
 and equality holds when $\alpha=1/2$. Thus we see that two cuts, one cut for the upper boundary line and one cut for the lower boundary line are sufficient, i.e. it suffices to set $k=1$. 
 \end{proof}

\subsection{A Partially Optimized Expression for $P_e$}
\label{sec:paropep}
From \eqref{eqn:bigprob} and \eqref{eqn:prob}, assuming an optimal partition of $\Rtwo$ by node 2,  and the fact that $W_1=0$ implies the error probability is zero, we have
\begin{equation}
P_e = \sum_{w_1=1}^m \sum_{z_1=0}^{N_{w_1}-1}\frac{(\lentwo)^2(|\dot{u}|+|\dot{l}|)}{4\lengthbabai\heightbabai}.
\end{equation}

Define $\gamma~:~\{0,1,\ldots,m\}\to \mathbb{R},$ by
\begin{equation}
\gamma_{w_1}=\frac{(|\dot{u}|+|\dot{l}|)}{4\heightbabai},
\end{equation}
where $\dot{u}$ and $\dot{l}$ are determined by the argument of $\gamma$, namely $w_1$.
Then
\begin{IEEEeqnarray}{rCl} 
 P_e & = & \sum_{w_1=1}^m \gamma_{w_1}\sum_{z_1=0}^{N_{w_1}-1}\frac{(\lentwo)^2}{\lengthbabai} \nonumber \\
& = &   \sum_{w_1=1}^m \underbrace{\gamma_{w_1}\Big(\sum_{z_1=0}^{N_{w_1}-1}\lentwo\frac{\lentwo}{\lenone} \Big)}_{\prob{\bm{X}\in\cE\mid\bm{X}\in \Rone}} \nonumber \\
& & \hspace{5.0cm} \times \frac{\lenone}{\lengthbabai}.
\label{eqn:cleanpe}
\end{IEEEeqnarray}

\subsection{Communication Rates}
\label{sec:commrates}
The communication rate for Stage-I is $H(U_2)+H(U_1|U_2)$ where $U_1, U_2$ are given in \eqref{eqn:Babai2D} and is calculated using
\begin{equation}
\ent{U_2}+\conent{U_1}{U_2}=\ent{\mynint{\frac{X_2}{|\rho \sin \theta|}}}+\ent{\mynint{X_1}}.
\end{equation}
Further, since $H(U_1,U_2|\Latpt_{np})=H(\Latpt_{np}|U_1,U_2)=0$, it follows that
\begin{equation}
\ent{U_2}+\conent{U_1}{U_2}=\ent{\Latpt_{np}}.
\end{equation}
We observe that the Stage-I communication rate is determined by the two-dimensional differential entropy of the source and the determinant of the lattice (see (39) in \cite{VSlSer:2001}). For this reason the focus of this paper is on the Stage-II communication rate, 
given by  $\ent{W_1,Z_1,Z_2|\Latpt_{np}}$.  Evaluation of the Stage-II communication rate is  simplified by the following independence result.
\begin{theorem}
$(W_1,Z_1,Z_2)$ is independent of $\Latpt_{np}$.
\end{theorem}
\begin{IEEEproof}
\begin{IEEEeqnarray}{rCl}
\lefteqn{\prob{W_1=w_1,Z_1=z_1,Z_2=z_2\mid \Latpt_{np}=\latpt} = } &   \nonumber \\  
  & = &  \frac{\prob{W_1=w_1,Z_1=z_1,Z_2=z_2, \Latpt_{np}=\latpt} }{\prob{\Latpt_{np}=\latpt} } \nonumber \\
& = & \frac{\prob{\bm{X}\in \Rthree+\latpt}}{\prob{\bm{X}\in \Bab+\latpt}} \nonumber \\
& = &  \frac{\vol{\Rthree}p_\latpt}{\vol{\Bab}p_\latpt} \nonumber \\
& = & \prob{W_1=w_1,Z_1=z_1,Z_2=z_2},
\end{IEEEeqnarray}
where the second step follows from the translation structure of the algorithm and the equivalence mentioned at the end of Sec.~\ref{sec:babvorgeom}
\end{IEEEproof}
Thus, by an application of the chain rule,  the Stage-II communication rate is given by
\begin{eqnarray}
\ent{W_1,Z_1,Z_2} & = & {\conent{Z_2}{W_1,Z_1}+\conent{Z_1}{W_1}+\ent{W_1}}, \nonumber \\
\end{eqnarray}
individual terms of which are calculated using
\begin{IEEEeqnarray}{lCl}\label{eqn:hzw}
\conent{Z_1}{W_1}  =  \sum_{w_1=0}^m\frac{\lenone}{\lengthbabai} \times \nonumber \\
 \underbrace{ \sum_{z_1=0}^{N_{w_1}-1} \frac{\lentwo}{\lenone}\log \frac{\lenone}{\lentwo}}_{\conent{Z_1}{W_1=w_1}},\nonumber \\
\ent{W_1} =  \sum_{w_1=0}^m\frac{\lenone}{\lengthbabai}\log \frac{\lengthbabai}{\lenone}, \nonumber \\
\conent{Z_2}{W_1,Z_1}  =  \sum_{w_1=0}^m\sum_{z_1=0}^{N_{w_1}-1}\frac{\lentwo}{\lengthbabai} \times \nonumber \\
\sum_{z_2=-1}^1\frac{\lenthree}{\heightbabai} \log\frac{\heightbabai}{\lenthree}.
\end{IEEEeqnarray}

Note that $\conent{Z_2}{W_1,Z_1}$ must be evaluated numerically, since it does not appear to have an explicit formula. When  the intervals $\Intone$, $w_1\neq 0$ are partitioned finely  by $Z_1$, it is possible to approximate this by an integral, which must also be evaluated numerically, see Appendix~\ref{sec:app3} for more details.

Note that $\conent{Z_2}{W_1,Z_1}\leq \prob{W_1\neq 0}\log 3$, regardless of how finely $Z_1$ partitions $\Bab^{(1)}$. On the other hand, $\conent{Z_1}{W_1}$ increases logarithmically with the size of the partition imposed by $Z_1$. For this reason, we will optimize the partition created by $Z_1$ while imposing a constraint on 
$\conent{Z_1}{W_1}$, even though $\conent{Z_2}{W_1,Z_1}$ exhibits a weak dependence on the partition at low rates.

\section{Optimization of the Partition}
\label{sec:perateoptim}
We minimize $P_e$ with respect to $\ent{W_1,Z_1,Z_2}$ in two steps. In the first step, we minimize $\prob{\bm{X}\in \Rone\bigcap \cE}$ subject to a constraint on  $\conent{Z_1}{W_1=w_1}$ (defined by the inner summation in \eqref{eqn:hzw}) over the interval lengths $\{\lentwo,~z_1=0,1,\ldots,N_{w_1}-1\}$. We then minimize $P_e$ subject to a constraint on $\conent{Z_1}{W_1}$. 
\begin{theorem}\label{thm:optim1}
The minimum of $\prob{\bm{X}\in\cE\mid \bm{X}\in \Rone}$ subject to $\conent{Z_1}{W_1=w_1}\leq \Rate_{w_1}$ can be expressed parametrically
in terms of $N_{w_1}$, by
\begin{eqnarray}\label{eqn:parametric}
\prob{\bm{X}\in\cE\mid \bm{X}\in \Rone} & = & \frac{\gamma_{w_1}\lenone}{N_{w_1}}, \nonumber \\
\Rate_{w_1} & = & \log N_{w_1},
\end{eqnarray}
$w_1=1,2,\ldots,m$  and is attained by partitioning $\Intone$ into $N_{w_1}$ bins of equal length, i.e. by setting $\lentwo=\lenone/N_{w_1}$, $z_1=0,1,\ldots,N_{w_1}-1$.
\end{theorem}
\begin{IEEEproof}
See Appendix~\ref{sec:app1}.
\end{IEEEproof}

The next result is a second optimization, in which the rates $\Rate_{w_1}$, $w_1=1,2,\ldots,m$, are allocated optimally. This results  in a compact limiting form for the error probability as the rate grows to infinity, as the statement of the following theorem will show. In the sequel, we denote $\prob{W_1=w}$ by $P_{W_1}(w)$.
\begin{theorem}\label{thm:optim2}
The minimum value of $P_e$ subject to $\conent{Z_1}{W_1}\leq \Rate$ has  limiting form
\begin{IEEEeqnarray}{lCl}\label{eqn:optimpe}
\lim_{\Rate\to \infty} e^{\tfrac{\Rate}{(1-P_{W_1}(0))}}P_e = \nonumber \\
(1-P_{W_1}(0)) \prod_{w_1=1}^m\left(\gamma_{w_1}\lenone\right)^{\frac{P_{W_1}(w_1)}{(1-P_{W_1}(0))}} =: \Xi_1(\theta),
\end{IEEEeqnarray}
which is attained by selecting $\Rate_{w_1}$ such that 
\begin{IEEEeqnarray}{lCl}\label{eqn:optimrw}
\lim_{\Rate\to \infty} \left(\Rate_{w_1}-\frac{\Rate}{(1-P_{W_1}(0))} \right)= \nonumber \\
\log\left(\frac{\gamma_{w_1}\lenone}{\prod_{w_1\neq 0}\left(\gamma_{w_1}\lenone \right)^{P_{W_1}(w_1)/(1-P_{W_1}(0))}} \right),
\end{IEEEeqnarray}
$w_1=1,\ldots,m-1,m$.
\end{theorem}
\begin{IEEEproof}
See Appendix~\ref{sec:app2}.
\end{IEEEproof}
\begin{remark}
\label{rem:sumrate}
Let $\Xi_2(\theta)=e^{\conent{Z_2}{W_1,Z_1}/(1-P_{W_1}(0))}$ and $\Xi_3(\theta)=e^{\ent{W_1}/(1-P_{W_1}(0))}$ and $\Xi(\theta)=\Xi_1(\theta)\Xi_2(\theta)\Xi_3(\theta)$. Then the minimum value of $P_e$ subject to $\ent{W_1,Z_1,Z_2}\leq \Rate_{II}$ has limiting form
\begin{equation}
\lim_{\Rate_{II}\to \infty} e^{\Rate_{II}/(1-P_{W_1}(0))}P_e=\Xi(\theta).
\label{eqn:optimrwnew}
\end{equation}
\end{remark}

\section{The Infinite Round Interactive Protocol}
\label{sec:infinitedec}

We now analyze the interactive model in which an infinite number of communication rounds are allowed in Stage-II. The protocol presented here is referred to as the bit-exchange protocol\footnote{Even though we were unaware of this at the time, it turns out that a similar bit-exchange protocol was suggested by Max Costa for a different problem~\cite{Orlitsky:1992}! Our observation that it results in zero error at finite rate appears to be new.}.

The observed vector $\bm{X}$ lies in $\Bab$ after subtraction of $\Latpt_{np}(\bm{X})$. The first round of communication, referred to as round-0,  is special. Rounds that follow all have a fixed protocol. At the conclusion of each round of the protocol, both nodes know that $\bm{X}$ lies in a specific rectangle. This rectangle is said to be \emph{error-free} if it lies entirely in a Voronoi cell, else it is said to be \emph{errored}.

\begin{figure}[h!]
\begin{center}
\includegraphics[width=2.5in]{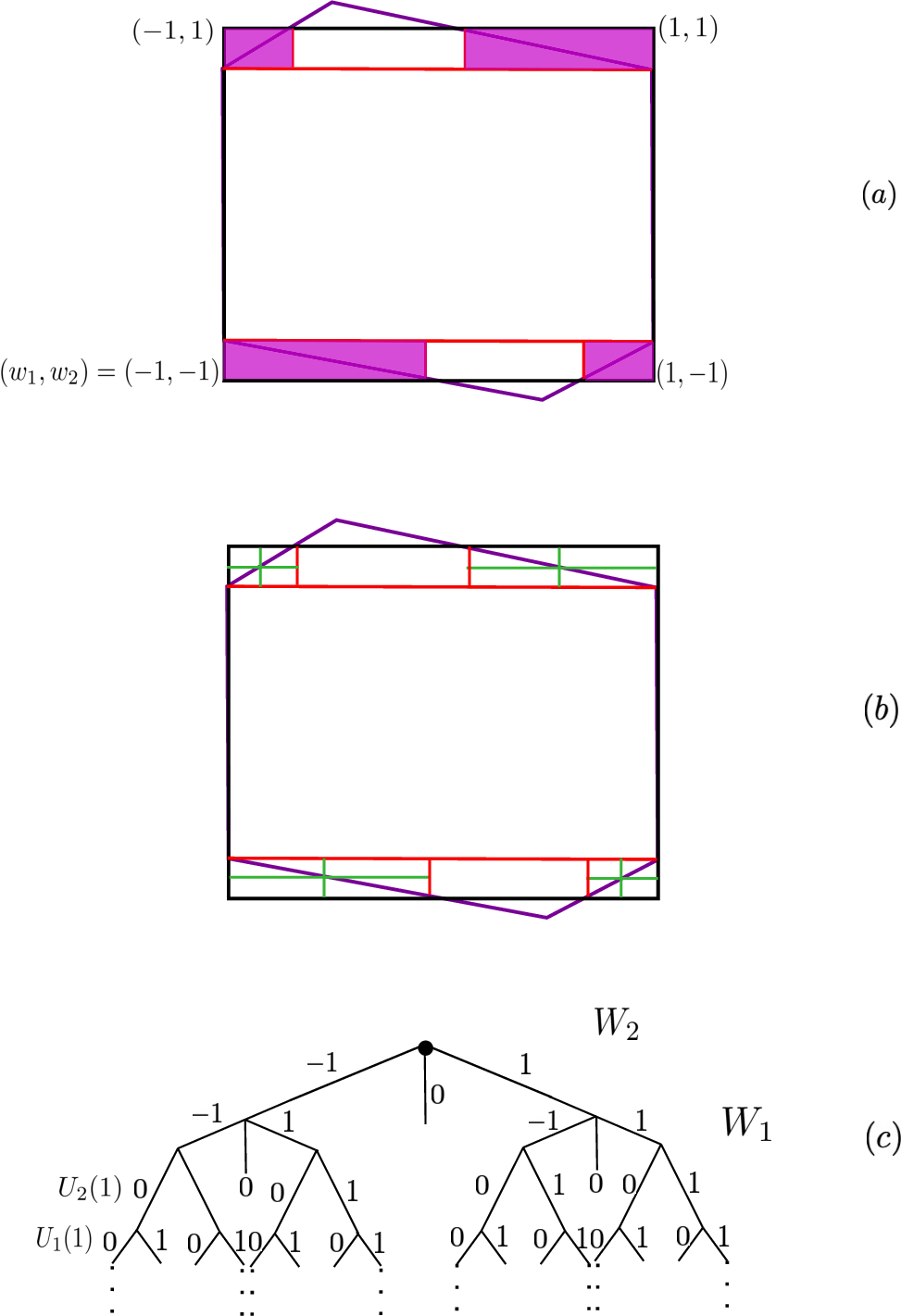}  
\caption{Partition created by the infinite round protocol (a) after one round (b) after two rounds. (c) Protocol tree for the infinite-round protocol. }
\label{fig:infint1}
\end{center}
\end{figure}

In round-0, node-2 communicates first and sends index $w_2$ to node-1. In effect, node-2
partitions $\Bab^{(2)}$ into three intervals,  
$\Intervaltwo_{-1}=[-\rho \sin \theta/2,-(\rho-\cos \theta)/(2\sin \theta)),$  $\Intervaltwo_0=(-(\rho-\cos \theta)/(2\sin \theta), (\rho-\cos \theta)/(2\sin \theta)],~ \text{and}~ \Intervaltwo_1=-\VInterval_{-1}$, and  random variable $W_2$ is the index of the interval in which  $X_2$ lies. In round-0, upon receiving $W_2$ and if  $W_2=1$, node-1 partitions  $\Bab^{(1)}$ into three intervals  $\Intervalone_{-1}=(-1/2,t_{-2}]$, $\Intervalone_0=(t_{-2},t_1]$ and $\Intervalone_1=(t_1,1/2]$, where $t_{-2}=(-1+\rho \cos \theta)/2$ and $t_1=\rho \cos \theta /2$. If $W_2=-1$, the support of $\Bab^{(1)}$ is partitioned into intervals $-\Intervalone_1,-\Intervalone_0, -\Intervalone_{-1}$. If $W_2=0$, no partitioning step is taken. Random variable $W_1$ describes which of the intervals $X_1$ lies. Let $Pr(W_2=i)=Q_i$, $i=-1,0,1$. Let $P_i=Pr(W_1=i|W_2=1)$, $i=-1,0,1$. Let $Q=(Q_0,Q_1,Q_2)$ and $P=(P_0,P_1,P_2)$. The partition of $\Bab$ after a single round is shown in Fig.~\ref{fig:infint1}. Of the seven rectangles three are error free and four are errored. Communication stops after round-$1$ if $\bm{X}$ lies in one of the error-free rectangles, else it continues.

At the conclusion of  the round-$0$ suppose $\bm{X}$ lies   in an errored rectangle  $[a,b)\times [c,d)$.   The rectangle is known to both nodes. Node-$i$ rescales $X_i$ as follows. 
\begin{eqnarray}
X_2  & \leftarrow &  \frac{X_2-c}{d-c}   \nonumber \\
X_1 &  \leftarrow & \left\{ \begin{array}{cc}  \frac{X_1-a}{b-a} & W_1\neq W_2 \vspace{0.1cm} \\ 
 \frac{X_1-b}{a-b} & W_1 = W_2. \end{array} \right.
\end{eqnarray}
Let $U_i(k)$ be the $k$th bit in the binary expansion of $X_i$, thus $X_1=U_1(1)/2+U_1(2)/2^2+\ldots$. In round-$k$, $k=1,2,\ldots$, assuming that the protocol has not stopped in the previous round, node-$i$ sends $U_i(k)$ to the other node. The protocol stops at round-$k$ if  $U_1(k)\neq U_2(k)$, else it moves to round-$(k+1)$. 
The partition of the Babai cell  after two rounds of communication is  shown in Fig.~\ref{fig:infint1} along with a tree representation of the protocol. 

Let $N({\bm X})$, $\Rate({\bm X})$  denote the number of rounds, and number of bits communicated, respectively, when the algorithm halts. Let $\bar{\Rate}=E[\Rate({\bm X})]$ and $\bar{N}=E[N({\bm X})]$ denote averages over ${\bm X}$.
\begin{theorem}
For the interactive model with unlimited rounds of communication, the Voronoi point $\latpt_v(\bm{X})$ is obtained, after a finite number of bits and rounds of communication, on average. Specifically, $\prob{\Latpt_\Pi(\bm{X})\neq \Latpt_{v}(\bm{X})}=0$, 
\begin{equation}\label{eqn:avrateinf}
\bar{\Rate}=\ent{Q}+(1-Q_0)\ent{P}+4(1-P_0)(1-Q_0), ~ \textnormal{and}~
\end{equation}
$$\bar{N}=1+2(1-P_0)(1-Q_0).$$
\end{theorem}
\begin{proof}
We assume that an optimum entropy code is used (thus if $W_2=0$, the codeword length is $\log_2(1/Q_0)$ bits). The term $\ent{Q}+(1-Q_0)\ent{P}$ in (\ref{eqn:avrateinf}) is the cost of resolving the round-1 partition. At the conclusion of round-1, $\bm{X}$ belongs to an errored region with probability $(1-P_0)(1-Q_0)$. The average number of bits transmitted thereafter is obtained by the following argument. At the conclusion of round $k$, $\bm{X}$ has been localized to an error-free rectangle if and only if $U_1(k)\neq U_2(k)$. Further, ${U_1(k)}$ and ${U_2(k)}$ are two independent sequences of independent  unbiased Bernoulli random variables.  The result follows immediately because the probability of stopping at round $k$ is $2^{-k}$, $k=1,2,\ldots$, given that it has continued past round $0$.
\end{proof}

\begin{remark}
This result has  interesting implications when viewed in the context of distributed classification problems. Suppose we have an optimum two-dimensional classifier with separating boundaries that are not axis aligned and also a suboptimal classifier with separating boundaries that are axis aligned, e.g. a $k$-$d$ tree. We expect the communication complexity of refining the approximate rectangular classifier to the optimum classifier to be finite.
\end{remark}

%
%
%

\section{Numerical Results and Discussion}
\label{sec:discussion}
\begin{figure}[htbp] 
   \centering
   \includegraphics[width=3.4in]{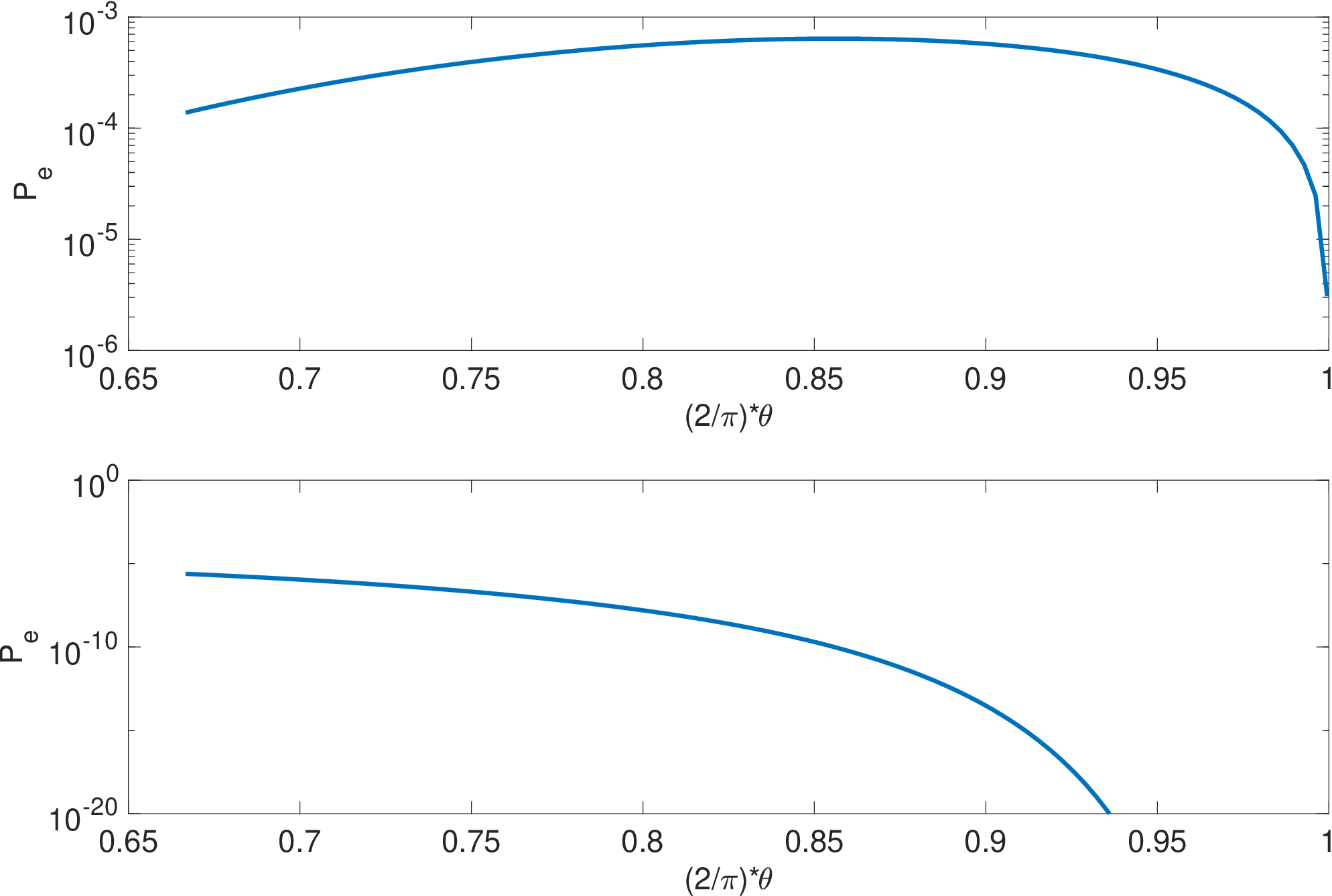} 
     \caption{Variation of $P_{e}$ with $\theta$, $\rho=1.0$,  for the single-round interactive model, 12 (top), 21 (bottom) with  $\ent{W_1,Z_1,Z_2}=4.0$ bits. $P_{e} <10^{-20}$ not shown in bottom plot.}
   \label{fig:thetavariation}
\end{figure}

Performance results for the two single-round interactive models (i.e. the $12$ and $21$ orders) as well as the infinite round interactive model are shown in Fig.~\ref{fig:thetavariation}, for $\rho=1$ and $\pi/3 < \theta < \pi/2$.  The case $\theta=\pi/3$ corresponds to the hexagonal lattice $A_2$ and $\theta=\pi/2$ corresponds to the square lattice ${Z}_2$. The additional calculations  required are in Appendix~\ref{sec:app3}. 

\begin{figure}[htbp] 
   \centering
   \includegraphics[width=3.4in]{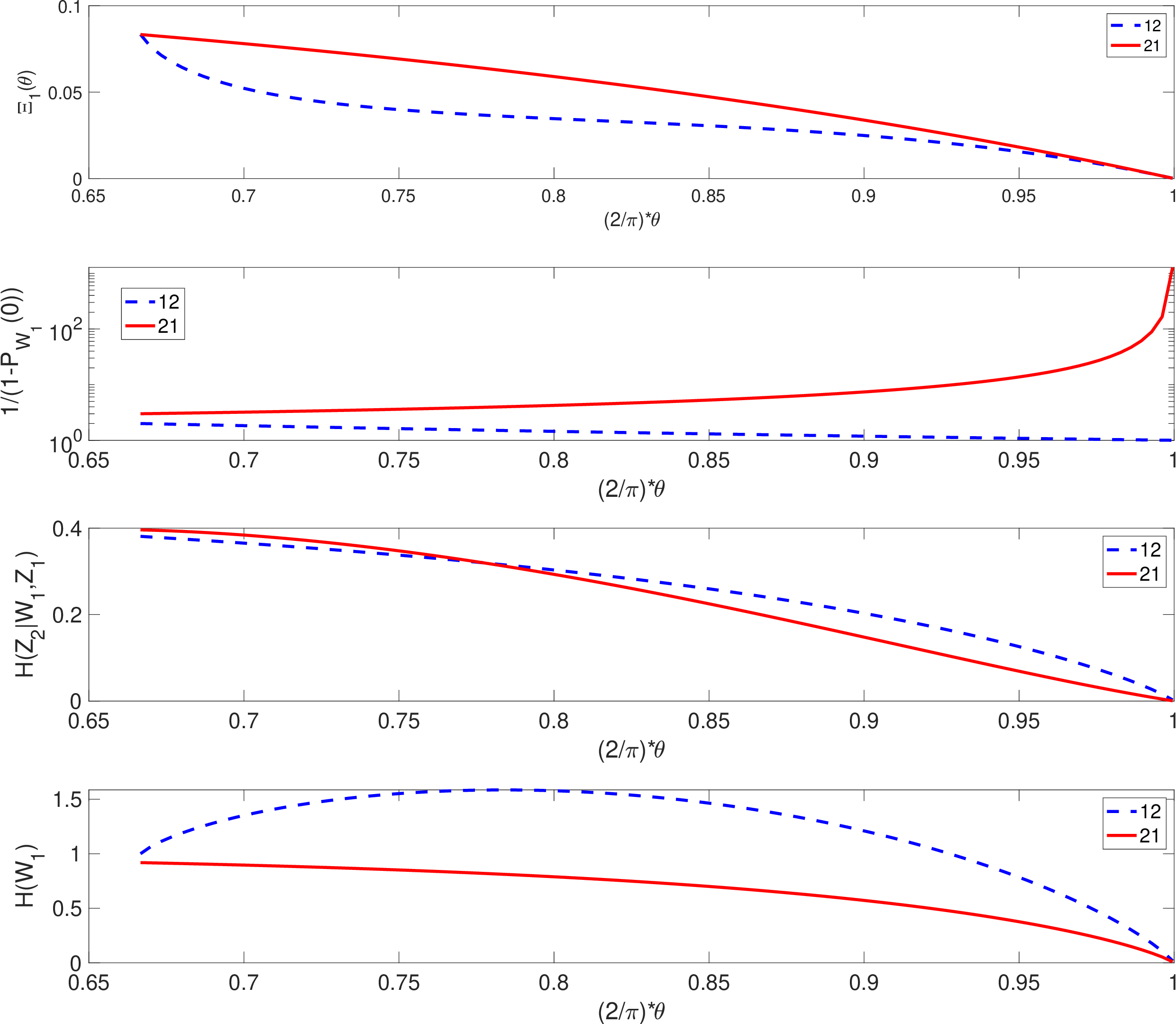} 
   \caption{Various terms that contribute to the error probability. $\Xi_1(\theta)$ is the limit in \eqref{eqn:optimpe}. $\rho=1.0$.}
   \label{fig:breakout}
\end{figure}

The error probability is computed for a rate $H(W_1,Z_1,Z_2)=4.0$ bits in Fig.~\ref{fig:thetavariation}. Fig.~\ref{fig:breakout} shows the variation of various terms that contribute to the error probability and to the rate. 

Note that the error probability  for the $21$ order is significantly smaller than for the $12$ order. It is also worth noting that $1/(1-P_{W_1}(0))$, the term that multiplies the rate, goes to infinity for the $21$ order but goes to zero for the $12$ order. This is because of  the structure of the Babai cell. For the $12$ order, the rectangle $\Rone$ with $w_1=1$ dominates as $\theta \to \pi$ whereas for the $21$ order, $\Ronezero$ dominates the Babai cell $\Bab$. 

\begin{figure}[htbp] 
   \centering
   \includegraphics[width=3.5in]{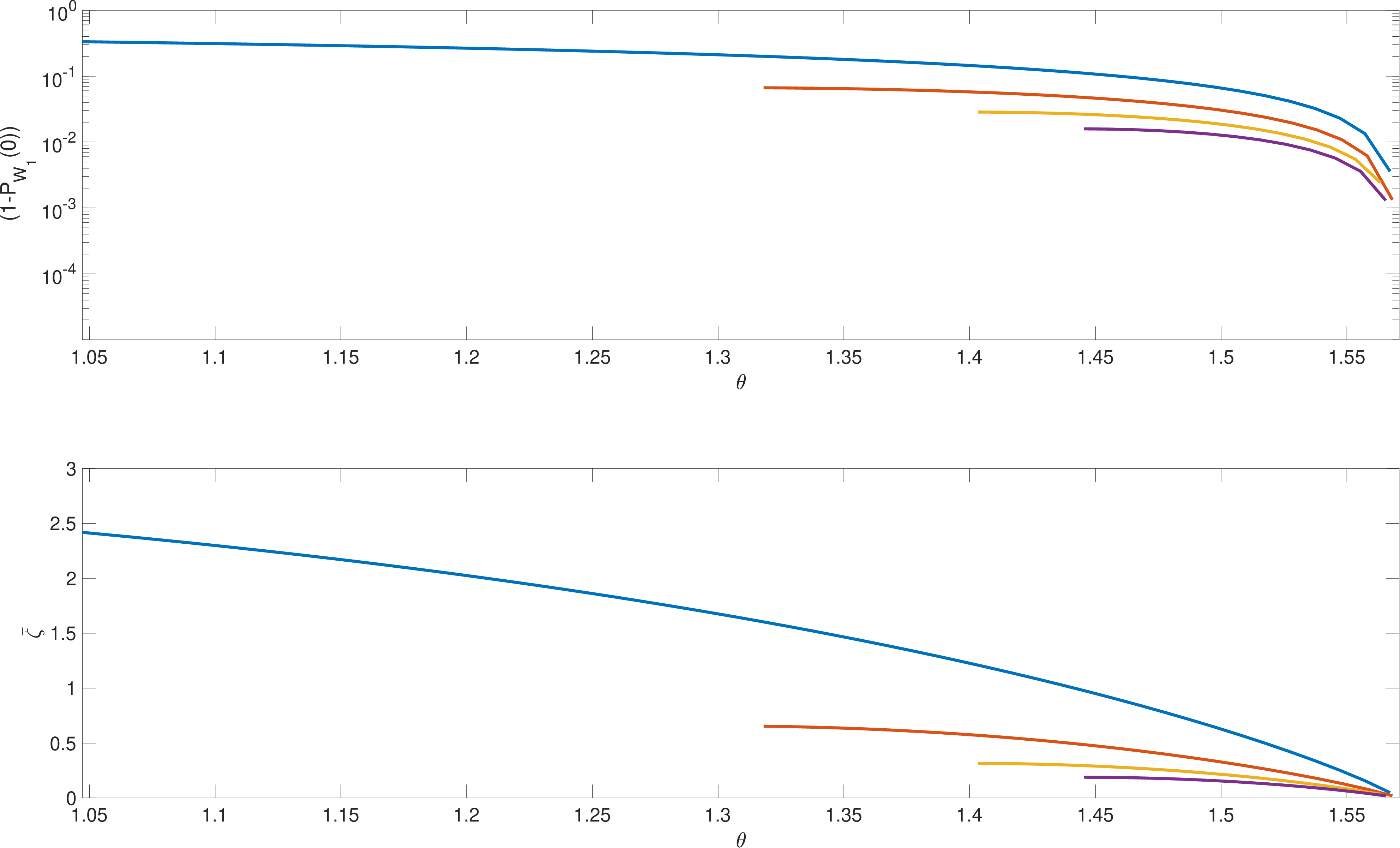} 
   \caption{Variation  of communication cost with respect to  $\rho$ and $\theta$. In each plot curves are for $\rho=1.0,2.0,3.0, 4.0$, resp.,  from top to bottom. Top plot: $21$ order for the single-round protocol, the quotient of the rate exponent $(1-P_{W_1}(0))$. bottom plot: communication rate for the infinite round protocol. }
   \label{fig:infround}
\end{figure}

Under the single-round interactive model  and considering the best ordering (which is $21$) the hexagonal lattice, with basis $\{(1,0), (1/2,\sqrt{3}/2)\}$---the  best lattice for quantization and coding in two dimensions~\cite{SPLAG}---has the highest communication cost, see top plot in Fig.~\ref{fig:infround}.  The large gap in performance at the same rate for the $12$ and $21$ sequences highlights the importance of selecting the  order in which nodes communicate. 

 The communication rate for achieving zero error probability for the infinite round protocol is shown in Fig.~\ref{fig:infround}, bottom plot. Under the infinite round interactive protocol, the hexagonal lattice  once again has the highest communication cost, with $\bar{\Rate}=2.42$ bits. 
 
\section{Generalization to dimension $n>2$}
\label{sec:highdim}
While the problem has been formulated for all dimensions $n\geq 2$, the study in this paper has been limited to $n=2$. This is for the following reasons. (i) $n=2$ this is the smallest dimension in which the shape of a Voronoi cell plays a role in the communication cost, (ii) the partition can be easily visualized, which helps develop useful intuition, (iii) a complete parameterization of lattices exists in two dimensions and is not available in higher dimensions; this will preclude as thorough a study in higher dimensions, (iv) the higher dimensional case can be reduced to the study of the two-dimensional case (work in progress), since for the single round results, a transmission from node $i$  essentially reduces the dimension of the problem for the remaining nodes by one. Eventually, through the use of a bounding argument, the problem is reduced to a series of two-dimensional problems.  Fully developing this approach requires a significant amount of added machinery which will be considered in a later submission, (v) for the infinite round case, new machinery may be needed, see, e.g.,~\cite{VVclassifier:2019}, (vi) quantization techniques in networks with two nodes are of interest and have been  considered recently in classification problems \cite{DZ2021}. 
\section{Summary and Conclusions}
\label{sec:summary}
We have considered the problem of interactively computing the  nearest lattice point for a lattice in two dimensions. A two-party model of communication is assumed and expressions for the error probability have been obtained for a single round of communication (i.e. two messages). We have also considered an unbounded number of rounds of communication and shown that it is possible to achieve zero probability of error with a finite number of bits exchanged on average. Our results indicate that lattices which are better for quantization or for communication have a higher communication cost.


\section{Acknowledgement}
We thank the reviewers and the AE for their feedback and suggestions.
\appendices	
\section{Proof of Theorem~\ref{thm:optim1}}	
\label{sec:app1}

\begin{IEEEproof}
From \eqref{eqn:cleanpe} we have
\begin{IEEEeqnarray}{lCl}\label{eqn:firstoptim}
\lefteqn{\prob{\bm{X}\in \cE \mid \bm{X}\in \Rone}=} & \nonumber \\
& = & \gamma_{w_1}\sum_{z_1=0}^{N_{w_1}-1}\lentwo \frac{\lentwo}{\lenone} \nonumber \\
& =  & \gamma_{w_1}\sum_{z_1=0}^{N_{w_1}-1}\exp\left\{\log(\lentwo\right\} \frac{\lentwo}{\lenone} \nonumber \\
& \stackrel{(a)}{\geq} &  \gamma_{w_1} \exp\left\{ \sum_{z_1=0}^{N_{w_1}-1} \log(\lentwo)\frac{\lentwo}{\lenone} \right\} \nonumber \\
& = &  \gamma_{w_1} \exp\left\{ \sum_{z_1=0}^{N_{w_1}-1} \left(\log\left(\frac{\lentwo}{\lenone}\right)+\log \lenone\right) \right. \times \nonumber \\
&  & \hspace{5.0cm} \left. \frac{\lentwo}{\lenone} \right\} \nonumber \\
& = &  \gamma_{w_1}\lenone \exp\left\{-\Rate_{w_1} \right\},
\end{IEEEeqnarray}
where in (a) we have used Jensen's inequality and equality holds when $\lentwo$ is a constant. Since $\sum_{z_1=0}^{N_{w_1}-1}\lentwo=\lenone$, it follows that $\lentwo=\lenone/N_{w_1}$, $z_1=0,1,\ldots,N_{w_1}-1$. Thus, the optimum error probability and communication rate tradeoff can be expressed parametrically by \eqref{eqn:parametric}.
\end{IEEEproof}

\section{Proof of Theorem~\ref{thm:optim2}}
\label{sec:app2}
Our objective is to select  $N_{w_1}\geq 0$,  $N_{w_1} \in \mathbb{Z}$, $w_1=1,2,\ldots,m$, so as to minimize the error probability subject to a rate constraint. We will consider a relaxation of the problem, to $N_{w_1} \in \mathbb{R}$, since in the limit as the rate goes to infinity, the error introduced is negligible.
Recall the notation $P_{W_1}(w)=\prob{W_1=w}$. 
\begin{IEEEproof}
From \eqref{eqn:firstoptim}
\begin{IEEEeqnarray}{rCl}
\lefteqn{\prob{\bm{X}\in \cE} = }&   \nonumber \\
& = & \sum_{w_1=1}^m\gamma_{w_1}\lenone\exp(-\Rate_{w_1})P_{W_1}(w_1) \nonumber \\
& = & (1-P_{W_1}(0))\sum_{w_1=1}^m \exp\{\log(\gamma_{w_1}\lenone\exp(-\Rate_{w_1}))\} \nonumber \\ & & \hspace{4.5cm} \times ~ \frac{P_{W_1}(w_1)}{(1-P_{W_1}(0))} \nonumber \\
&\stackrel{(a)}{ \geq} & (1-P_{W_1}(0))\exp\left\{ \sum_{w_1=1}^m\log(\gamma_{w_1}\lenone\exp(-\Rate_{w_1})) \right. \nonumber \\ & & \hspace{4.5cm} \left. \times~ \frac{P_{W_1}(w_1)}{(1-P_{W_1}(0))} \right\}, 
\end{IEEEeqnarray}
where (a) follows from Jensen's inequality, and equality holds when $\gamma_{w_1}\lenone\exp(-\Rate_{w_1})=C$ is a constant independent of $w_1$ for $w_1\neq 0$. Since $$\sum_{w_1=1}^m\Rate_{w_1}\prob{W_1=w_1}=\Rate,$$ it follows that
$$
C=\prod_{w_1=1}^{m}\left(\gamma_{w1}\lenone\right)^{\frac{P_{W_1}(w_1)}{(1-P_{W_1}(0))}}\exp\left( -\frac{\Rate}{(1-P_{W_1}(0))}\right).
$$
The final results \eqref{eqn:optimpe} and \eqref{eqn:optimrw} follow from
$$
\prob{\bm{X}\in \cE}=(1-P_{W_1}(0)) C,$$
and
$$
\Rate_{w_1}=\log\left( \frac{\gamma_{w_1}\lenone}{C}\right),
$$
respectively.
\end{IEEEproof}

\section{Calculation Details for Single-Round Interactive Communication}
\label{sec:app3}

Two orders are possible $12$ and $21$. In each case, we need to determine $m$, $\gamma_{w_1}$, $\lenone$ and $P_{W_1}(w_1)$, $w_1=0,1,\ldots,m-1,m$. First consider the $12$ order of communication. 
The Voronoi cell $\Vor$ and  Babai cell $\Bab$ are shown with all the significant boundary points and intervals   in Fig.~\ref{fig:detailedGeom}. $\Bab$ has height $\heightbabai=\rho \sin \theta$ and length $\lengthbabai=1$ and  $m=3$. $W_1=2$ identifies $[-1/2,-(1-\rho \cos \theta)/2]\bigcup [(1-\rho \cos \theta)/2,1/2]$,  $W_1=1$ identifies $[-(1-\rho \cos \theta)/2,-\rho \cos \theta/2]\bigcup [\rho \cos \theta/2,(1-\rho \cos \theta)/2]$  and $W_1=0$ identifies $[-\rho \cos \theta/2,\rho \cos \theta/2]$.

The lengths are 
\begin{equation}\label{eqn:Lw1}
\lenone=\left\{\begin{array}{cc} \rho \cos \theta, & w_1=0, \\
1-2\rho \cos \theta, & w_1=1, \\
\rho \cos \theta, & w_1=2.
 \end{array} \right.
\end{equation}
Thus
\begin{equation}\label{eqn:gammaw1}
\gamma_{w_1}=\left\{ \begin{array}{cc} 
0, &  w_1=0,  \\
\cos \theta/ (4 \rho \sin^2 \theta), & w_1=1 \\
1/(4\rho^2 \sin^2 \theta), & w_1=2,
\end{array}\right.
\end{equation}
and 
\begin{equation}\label{eqn:pw1}
P_{W_1}(w_1)=\frac{\lenone}{\lengthbabai},~w_1=0,1,\ldots,m-1,m.
\end{equation}
For  the $12$  ordering, we approximate $\conent{Z_2}{W_1,Z_1}$ by an integral as follows.
For $x_1\in\Bab^{(1)}$, i.e. for $-1/2\leq x_1 \leq 1/2$, define the probability vector 
$$\bm{Q}(x_1)=\begin{pmatrix} \tfrac{\rho \sin \theta/2-u(x_1)}{\heightbabai}, & \tfrac{u(x_1)-l(x_1)}{\heightbabai}, & \tfrac{l(x_1)-(-\rho \sin \theta/2)}{\heightbabai}\end{pmatrix}$$
$x_2=\rho \sin \theta/2$ is the  upper boundary of $\Bab$.  Let
$\mathbb{G}(x_1)$ denote the entropy of $\bm{Q}(x_1)$. Then we estimate $\conent{Z_2}{W_1,Z_1}$ by
$$
\int_{-1/2}^{1/2} \mathbb{G}(x_1)dx_1.
$$


 The analysis of the   $21$ order of communication is similar and \eqref{eqn:optimpe} can be reused, by flipping, i.e. interchanging,  $x_1$ and $x_2$ in Fig.~\ref{fig:detailedGeom}. 
 Several points are worth noting. The random variable $W_1$ takes on only two values now, unlike in the $12$ order, where it took on three values. Specifically, $W_1=0$ corresponds to the interval $[-(\rho-\cos \theta)/(2\sin \theta), (\rho-\cos \theta)/(2\sin \theta))$ and $W_1=1$ corresponds to its complement in $\Bab^{(1)}$. The interval lengths are 
\begin{equation}
\lenone=\left\{\begin{array}{cc} \frac{(\rho-\cos \theta)}{\sin \theta}, & w_1=0, \\
\frac{\cos \theta (1-\rho \cos \theta)}{\sin \theta}, & w_1=1. 
\end{array} \right.
\end{equation}
Thus 
\begin{equation}
P_{W_1}(w_1)=\left\{\begin{array}{cc}
\frac{\cos \theta (1-\rho \cos \theta)}{\rho \sin^2 \theta}, & w_1=1 \nonumber \\
\frac{\rho-\cos \theta}{\rho \sin^2 \theta}, & w_1=0.
\end{array} \right.
\end{equation} 
Also
\begin{equation}
\gamma_{w_1}=\left\{ \begin{array}{cc} 
0, &  w_1=0, \nonumber \\
\frac{\sin \theta}{4 \cos \theta(1-\rho \cos \theta)}, & w_1=1.\nonumber \\
\end{array}\right.
\end{equation}
The optimal error probability for the $21$ order is determined by plugging values into \eqref{eqn:optimrw} and has the simple form
\begin{IEEEeqnarray*}{lCl}
\lim_{\Rate\to \infty}\prob{\bm{X}\in \cE}\exp{\left\{\frac{\Rate\rho \sin^2 \theta}{\cos \theta (1-\rho \cos \theta)}\right\}} =  & \\
& \hspace{-0.5cm} \frac{\cos \theta (1-\rho \cos \theta)}{4 \rho \sin^2\theta}.
\end{IEEEeqnarray*}


A similar approximation as used for the $12$ ordering, is used for estimating $\conent{Z_2}{W_1,Z_1,\Latpt_{np}}$ for the $21$.

\end{document}